%% file: main.tex
\documentclass[sigconf,nonacm]{acmart}

\usepackage{url}

\usepackage{float}

\usepackage{todonotes}

\usepackage{graphicx}
\usepackage{pgfplots}

\usepackage[ruled,vlined,linesnumbered]{algorithm2e}

\usepackage[english]{babel}
\usepackage{amsthm}

\newtheorem{theorem}{Theorem}[section]

\usepackage{wrapfig}

\usepackage{multicol}

\usepackage{svg}

\usepackage{amsmath}

\usepackage{comment}

\usepackage[shortlabels]{enumitem}
\setlist{nolistsep}
\setlist{nosep}

\usepackage{boxedminipage}

\newcommand{\fakesub}[1]{{\noindent\textcolor{black}{\textbf{#1}.}}}

\newcommand{\fakesubb}[1]{{\noindent\textcolor{black}{\textbf{#1}}}}

\makeatletter
\newcommand\resetstackedplots{
\makeatletter
\pgfplots@stacked@isfirstplottrue
\makeatother
\addplot [forget plot,draw=none] coordinates{(1,0) (2,0) (3,0)};
}
\makeatother

\usepackage{subcaption}
\usepackage[labelformat=simple]{subcaption}

\pgfplotsset{compat=1.17}

\usepackage{listings}
\usepackage{color}

\definecolor{dkgreen}{rgb}{0,0.6,0}
\definecolor{gray}{rgb}{0.5,0.5,0.5}
\definecolor{mauve}{rgb}{0.58,0,0.82}
\definecolor{aureolin}{rgb}{0.99, 0.93, 0.0}
\definecolor{bananayellow}{rgb}{1.0, 0.88, 0.21}
\definecolor{canaryyellow}{rgb}{1.0, 0.94, 0.0}
\definecolor{daffodil}{rgb}{1.0, 1.0, 0.19}
\definecolor{electricyellow}{rgb}{1.0, 1.0, 0.0}

\lstdefinestyle{myJava}
{ frame=tbrl,
  language=Java,
  numbers = left,
  stepnumber=1,
  showstringspaces=false,
  columns=flexible,
  basicstyle={\footnotesize\ttfamily},
  numberstyle=\tiny\color{gray},
  keywordstyle=\color{blue},
  commentstyle=\color{dkgreen},
  stringstyle=\color{mauve},
  breaklines=true,
  breakatwhitespace=true,
  tabsize=3,
  morekeywords={var, record, alias, vms, dep, event, String, Mono, Optional, Future, List, JsonObject, Map, select, from, where, and, set,forEach, getAttribute, eq}, 
  moredelim=[is][\color{gray}]{|}{|},
}

\usepackage{multirow}

\usepackage{tabularx}
\newcolumntype{Y}{>{\centering\arraybackslash}X}

\AtBeginDocument{%
  }

\setcopyright{acmlicensed}
\copyrightyear{2018}
\acmYear{2018}
\acmDOI{XXXXXXX.XXXXXXX}
\acmConference[Conference acronym 'XX]{Make sure to enter the correct
  conference title from your rights confirmation email}{June 03--05,
  2018}{Woodstock, NY}

\acmISBN{978-1-4503-XXXX-X/2018/06}

\begin{document}

\newcommand{\name}{vMODB}

\title{\name: Unifying Event and Data Management for Distributed Asynchronous Applications}

\author{Rodrigo Laigner}
\orcid{0000-0003-2771-7477}
\affiliation{
  \institution{University of Copenhagen}
  \city{Copenhagen}
  \country{Denmark}
}
\email{rnl@di.ku.dk}

\author{Yongluan Zhou}
\orcid{0000-0002-7578-8117}
\affiliation{
  \institution{University of Copenhagen}
  \city{Copenhagen}
  \country{Denmark}
}
\email{zhou@di.ku.dk}

\input{sections/00_abstract}



\received{20 February 2007}
\received[revised]{12 March 2009}
\received[accepted]{5 June 2009}

\maketitle

\input{sections/01_intro}
\input{sections/02_background}
\input{sections/03_vms}
\input{sections/04_system}

\input{sections/05_protocols}
\input{sections/06_implementation}
\input{sections/07_experiments}
\input{sections/09_conclusion}

\begin{acks}
This project has received funding from the European Union's Horizon 2020 research and innovation programme under the Marie Skłodowska-Curie agreement No 801199 and Independent Research Fund Denmark grant No 9041-00368B.

We thank Marcos Antonio Vaz Salles for contributions during the early versions of this work.
\end{acks}

\balance

\bibliographystyle{ACM-Reference-Format}
\bibliography{main}

\newpage
\appendix

\input{sections/99_appendix}


\end{document}

%% file: sections/00_abstract.tex
\begin{abstract}
Event-driven microservice architecture (EDMA) has emerged as a crucial architectural pattern for scalable cloud applications. In typical EDMAs, database systems are relegated to isolated storage engines for individual components, blind to cross-component transactions, while messaging systems are unaware of each component's application state. Consequently, EDMAs impose a severe trade-off: developers must either sacrifice strong data consistency and integrity or manually manage complex distributed coordination. To address this challenge, we design \name, a distributed framework that offers a better trade-off and enables developers to build highly consistent and scalable cloud applications without compromising the benefits of EDMA. The core contribution of \name\ lies in the co-design of a programming abstraction and the underlying specialized system. We propose Virtual Micro Service (VMS), a novel programming model that provides familiar Object-Relational Mapping (ORM) and meta-programming constructs for specifying the data model, constraints, concurrency, and dependencies, making application semantics visible to the system. \name\ leverages semantic visibility to enforce ACID properties by transparently unifying event logs and state management, relieving developers from the burden of ensuring cross-component data consistency and integrity. Thanks to full-stack system optimizations enabled by our co-design, experiments using two benchmarks show that \name\ outperforms a widely adopted state-of-the-art competing framework that only offers eventual consistency by up to 3x. 
\end{abstract}

%% file: sections/01_intro.tex
\section{Introduction}
\label{sec:intro}

Event-driven microservice architecture (EDMA) has emerged as a crucial architectural pattern for designing scalable cloud applications. EDMA encourages designing applications as independent components, each encapsulating a private state, communicating via asynchronous event messages. 
An event in EDMA serves the purpose of exchanging data, triggering remote
computations, and logging application behavior. 

Event producers and consumers in the system are oblivious to each other, promoting loose coupling between them~\cite{tanenbaum:2016}. The loose coupling of components boosts the independence of component development, allowing different teams to implement and evolve their components with little coordination, 
and brings about many technical benefits, including fine-grained resource provisioning, elastic scaling of individual components, adaptable event processing rate, etc.~\cite{bellemare2020building}. 


The dominant practice for EDMAs employs traditional web application frameworks such as Spring~\cite{spring}, ASP.NET~\cite{aspnet}, and Node.js~\cite{nodejs} to implement application components while relying on messaging systems to enable asynchronous communication~\cite{tanenbaum:2016}. Typically, EDMA follows design patterns resembling the BASE model~\cite{base} and SAGAS~\cite{sagas}, falling prey to the dangers of eventual consistency~\cite{eventual_consistency_today}. Recent studies~\cite{vldb2021,laigner2024benchmark,synapse} and reports~\cite{uber_microservice,uber_proxy,uber_money,uber_payment,nubank,nubank-arc,netflix_eda,netflix_delta,airbnb_integrity,airbnb_double,wix_pitfalls,wix,jet,b2w_query,ifood,podium} indicate that these practices pose developers significant challenges in maintaining data integrity and ensuring transactional properties. 

Being able to execute ACID transactions across multiple components in EDMAs would eliminate most of the reported challenges~\cite{vldb2021}. The technical reasons that EDMA gives up ACID are manifold. First, event messaging and log processing systems such as Kafka~\cite{kafka} are unaware of the progress of transaction execution and the state update operations at the individual components. Therefore, the best achievable outcome is eventual consistency, by retrying message delivery and eliminating the impact of duplicate messages~\cite{base,olep}. This task, per se, is complicated in an asynchronous system~\cite{deduplication}. 


Second, the database systems managing the states of individual components are oblivious to how components exchange events to complete a transaction spanning multiple components~\cite{data_outside}. In other words, they are unaware of the global structure of multi-component transactions or the data contained in the events representing the intermediate transaction states. In addition, a database system typically does not allow multiple asynchronous components to interact with the same database within the context of a single ACID transaction. Therefore, database systems cannot achieve ACID properties~\cite{berstein87} for multi-component transactions on their own.

Third, existing distributed commit frameworks (e.g., via OpenXA \cite{specification1991distributed}) require writing commit protocol phases explicitly in the application code. Besides being a challenging task for developers, this method deviates from the asynchronous event abstraction and breaks the desired state encapsulation of EDMAs. Hence, they are rarely employed in practice~\cite{vldb2021}.

To address the conundrum, we present a holistic system architectural shift. Departing from the status quo, we propose to unify event log and state management and make the system aware of the global structure of transactions, the progress of event processing, and the internal state of applications. The key enabler of this vision is a novel \emph{programming abstraction}. Unlike current solutions that treat EDMA components as opaque black boxes, our abstraction flips this paradigm by capturing deep semantic metadata, including dependencies, data constraints, and event flows, making the asynchronous logic visible to the underlying system. 
This visibility enables the system to transition from a passive data store to an active coordinator that schedules event processing to guarantee ACID properties and maintain data integrity, relieving developers' burdens while preserving the key properties of EDMA.

In this paper, we present \name, a novel distributed framework for building scalable and consistent asynchronous applications. The key contribution of \name\ does not stem from a specific algorithm or protocol, but rather from the co-design of a programming abstraction together with the underlying system execution and optimization. Our contributions are summarized as follows:


\begin{itemize}[left=0pt]
    \item We propose Virtual Micro Service (VMS), a programming model that allows developers to specify their application components, including their relational data model, data constraints and dependencies, and input and output event logs. It is built on the object-relational mapping (ORM) abstraction and meta-programming, both prevalent in industrial practice.  
    Developers can develop individual application components independently through an ordinary programming language by using the \name\ SDK, which provides the abstraction of data management and event-driven execution ($\S$~\ref{sec:prog_model}).
 

   \item Departing from conventional EDMA, where the individual components control when to process the incoming events, \name\ employs the Inversion of Control (IoC) principle~\cite{ioc}, where a scheduling service and the VMSes collaborate to control when to execute the VMSes' functions involved in a multi-VMS transaction to ensure transaction guarantees. The VMS abstraction enables \name\ to leverage the application semantics to minimize coordination and optimize concurrent execution, enabling various performance optimizations, including parallelization of VMS functions, deterministic transaction scheduling, and efficient commit and logging ($\S$~\ref{sec:architecture} \& $\S$~\ref{sec:protocols}).

   \item By unifying event and data management and leveraging the VMS abstraction, \name\ optimizes the entire system stack across I/O, caching, data processing, and task execution. Notably, \name\ contains a purpose-built multi-version database that is inherently aware of the event scheduling order via the VMS abstraction. As a result, it can mask in-progress writes and support consistent queries over VMS states and events, providing a coherent snapshot of the application state at any point in time ($\S$~\ref{sec:implementation}).
   
   \item Extensive experiments using two benchmarks show that, with the aforementioned co-design optimizations, \name\ outperforms state-of-the-art distributed frameworks that only offer eventual consistency by >3x, all while providing full ACID properties. Furthermore, \name\ demonstrates high scalability in TPC-C, proving its versatility ($\S$~\ref{sec:evaluation}). 
\end{itemize}

%% file: sections/02_background.tex
\section{Background and Motivation}
\label{sec:background}

\subsection{ORM-Based Transactional Applications}
The dominant design in applications that process transactions and serve queries to many users at low latency usually relies on relational databases that are abstracted away through object-relational mapping (ORM) facilities. 
ORM frameworks such as Hibernate~\cite{hibernate} map between tables in relational databases and application objects in advanced programming languages, and offer meta-programming features, such as annotations, to allow developers to express relational constraints (e.g., primary and foreign keys) through application objects and transactional properties, such as expressing whether operations in a block of code (e.g., a method) must be executed as a read-only or read-write transaction.

\vspace{-1ex}
\subsection{Event-Driven Microservice Architectures}

With the advent of cloud computing, practitioners realized that packaging application components as independent deployment units is critical to optimizing costs and enhancing application maintainability, evolvability, and scalability. This insight led to the rise of microservice architectures, where an application is broken down into small building blocks called microservices, each with specific functionalities and encapsulated states. Each microservice runs independently, allowing for tailored computational resources and scalability strategies, which can be managed separately by different teams in an organization, allowing for greater flexibility in the cloud while not breaking the ORM-based design of individual microservices. 

Despite the independent nature, microservices often exhibit data and functionality dependencies that necessitate coordination~\cite{vldb2021}. To achieve loose coupling while supporting such dependencies, microservice architectures increasingly opt for asynchronous, event-based communication. 
EDMAs manage microservice states following principles akin to BASE~\cite{base}, where transactions are modeled as workflows with asynchronous steps, each accomplished by a microservice. The availability of a microservice is not affected by the failure of the others. Microservices achieve eventual consistency through persistent asynchronous communication.
However, EDMA sacrifices the following data management features to achieve the aforementioned desired properties:

\begin{enumerate}[(i),wide,nosep]
\item \textbf{ACID transactions.} 
To adhere to state encapsulation and the asynchronous communication abstraction, EDMAs largely abandon distributed commit protocols~\cite{vldb2021}.

\item \textbf{Data invariant enforcement.} 
Due to state encapsulation, EDMAs rely on the application layer to enforce data invariants across microservices, e.g., foreign key constraints~\cite{base}.


\item \textbf{Query processing consistency.} 
Weak transaction isolation and event processing guarantees (e.g., at least once) make it difficult to 
ensure consistent query results.

\item \textbf{Data replication semantics.} 
Data caching and materialized views are critical mechanisms to improve performance in EDMAs~\cite{synapse}. However, enforcing data consistency must be done at the application layer~\cite{vldb2021}.


\end{enumerate}


\subsection{Related Work}
The prevalent practice for EDMA is to adopt a composite platform~\cite{vldb2021,synapse}, consisting of a web application framework (e.g., Spring) with an ORM framework (e.g., Hibernate) with a database system per microservice, and a message queue or event log system interconnecting all microservices (e.g., Kafka and Redis). To avoid loss or duplication of events and achieve application correctness, applications need to explicitly manage event offsets coherently with their executions.

Dapr is another popular framework for building EDMAs~\cite{dapr_github} and widely used in industry settings~\cite{dapr}. Dapr has its own built-in ORM framework and offers abstract key-value storage and event communication APIs, which rely on the support of external database systems and event communication systems. Therefore, it is basically a composite platform. It offers at-least-once event delivery by transparently managing the event offsets, relieving the development burden to a certain degree. Synapse~\cite{synapse} and MuCache~\cite{mucache} are microservice frameworks for consistent data replication and caching, respectively. However, as Dapr, they do not support ACID transactions across components.

Orleans~\cite{bykov2010orleans} is a framework that facilitates the development of distributed applications through the virtual actor model~\cite{bernstein2016orleans}. To support transactional applications, Orleans has recently introduced a transaction  API~\cite{orleans_transaction}. However, the Orleans Streams API, which enables decoupling actors in namespace and time, is not interoperable with the transaction API. Additionally, the model of fine-grained actors can impose challenges on porting traditional transaction applications~\cite{WangRBSMZ19,orleans_best_practices}.

\begin{table*}
\centering
\caption{Comparison of existing solutions for EDMAs in the cloud}
\vspace{-2ex}
\footnotesize
\begin{tabularx}{\linewidth}{|p{3cm}|X|X|X|X|p{1cm}|X|}
\hline
\textbf{Solution} & \textbf{Event-based comm.} 
& \textbf{Component Resource Isolation} 
& \textbf{State Encapsulation} & \textbf{Consistent Data Processing} & \textbf{ACID} & \textbf{Dev. Burden for Correctness} \\
\hline
Orleans Transactions & - & - & \checkmark & - & \checkmark & Low \\
\hline
Orleans Streams & \checkmark & - & \checkmark & - & - & High \\
\hline
Dapr & \checkmark & \checkmark & - & - & - & High \\
\hline
Other composite platforms& \checkmark & \checkmark & \checkmark & - & - & High \\
\hline
\textbf{\name} & \checkmark & \checkmark & \checkmark & \checkmark & \checkmark & Low \\
\hline
\end{tabularx}
 \vspace{-1ex}
\label{tab:comparison}
\vspace{-1ex}
\end{table*}

Some database systems, such as PostgreSQL~\cite{postgresql_pubsub}, offer asynchronous persistent messaging through pub/sub APIs. However, applications typically interact with DBMS via opaque libraries such as JDBC, which disallow ACID transactions across multiple asynchronous application components. 



Function as a Service (FaaS) systems, such as Statefun~\cite{statefun} and Netherite~\cite{netherite}, offer stronger execution guarantees; however, they do not offer multi-component ACID. Similar to Orleans~\cite{bykov2010orleans}, recently proposed transactional frameworks such as Snapper~\cite{snapper}, Styx~\cite{styx}, and Boki~\cite{boki}, offer programming models that force partitioning the application into fine-grained objects with single-thread access following the actor model~\cite{agha_actors_1986} and lack relational data models, data constraints, and query processing, which can be restrictive for a broad range of data-intensive applications~\cite{orleans_best_practices}. 

Distributed frameworks like Ambrosia~\cite{ambrosia} and Darq~\cite{darq} offer strong execution guarantees to application components but are oblivious to transactional data management.
 \vspace{-1ex}
\subsection{Addressing the "Elephant in the room"}
In Table~\ref{tab:comparison}, we compare popular platforms for EDMAs in the cloud.
The columns explore sought-after EDMA principles,
consistent data processing, which encompasses data invariant enforcement, event and data processing correctness, and the
development burden on achieving application correctness
while integrating and deploying multiple components.
Table~\ref{tab:comparison} shows that state-of-the-art platforms prioritize the core EDMA principles, namely, 
\textbf{asynchronous event-based communication},
\textbf{state encapsulation},
and \textbf{component-level resource isolation}, over the data management features, resulting in high development burdens in achieving application safety properties. Concerns over the data management trade-offs associated with EDMAs have recently sparked widespread debates about reverting to the monolithic design ~\cite{mendoncca2021monolith}. 

In this work, we argue that the monolithic architecture and the existing EDMA solutions only represent two extremes in the whole architectural spectrum, and there exist other models that can offer a better trade-off to practitioners yearning for both the EDMA properties and advanced data management features. More specifically, we intend to address the following key research question:
\textbf{Is it possible to achieve transactional data management and strong data consistency properties while withholding the core EDMA principles?}


The fundamental challenge of supporting advanced data management features in EDMA lies in the fact that there is data both on the outside~\cite{data_outside}, in the form of event exchanges, and on the inside, via microservices' encapsulated local states. 
We posit that unifying event and data management allows a system to be aware of the global structure of transactions, event processing progress, and the state of applications. Thereby, the system can schedule the processing of events to achieve advanced data management requirements without compromising the envisioned benefits of EDMA.

%% file: sections/03_vms.tex
\section{Programming Model}
\label{sec:prog_model}



To achieve advanced data management features in an EDMA system and reduce the burden on developers~\cite{vldb2021}, we need a programming model that provides the right abstraction level, allowing developers to build EDMA applications without managing concerns outside the application logic, such as network, transaction management, query processing, data, and event management.
We propose the virtual micro service (VMS) model, which offers key constructs for advanced data and event management, multi-microservice ACID transactions, and intra-microservice concurrency while allowing developers to specify the application functions of microservices and their communication through the event log abstraction. 

\subsection{Virtual Micro Services (VMS)}
\label{subsec:vms}



A \textit{VMS} contains a state $S$, a set of application functions $F$, a map $MI$, mapping some input event logs to the functions in $F$, and a map $MO$, mapping a function in $F$ to an output event log. Each event in the input event logs will trigger each of the functions it is mapped to by $MI$ exactly once. Each function $f$ in $F$ can be mapped by $MO$ to either one event log or none as its event output log. $f$ is an atomic set of actions over $S$ and its output event log specified by $MO$. The output event log of a VMS function should be unique in the whole system. 
Besides, there should be no overlapping among event logs in $MI$ and $MO$ in a \textit{VMS}.

\textit{VMS} uses the relational model for managing states. The state $S$ of a VMS comprises multiple mutable relational tables adhering to a relational database schema. Thus, VMS supports the specification of relational database constraints on the schema, including primary and foreign key constraints, non-null attributes, value-based constraints, cascading deletes, sequences, and more. The set of data items in $S$ can be further divided into 2 non-overlapping subsets: (1) $S_N$ - native tables, managed by this \textit{VMS}. (2) $S_F$ - foreign tables, owned by another \textit{VMS}. $S_F$ can be accessed, but not modified by this \textit{VMS}. More specifically, $S_F$ is a replication of a subset of data items owned by another \textit{VMS}.

VMSes communicate asynchronously via event logs, which follow the EDMA principles~\cite{tanenbaum:2016} to decouple VMSes in time. Each event log is a uniquely identified, append-only sequence of immutable events~\cite{immutability}, each being a tuple following a specific event schema. A log supports two operations: (1) \textit{Read}, retrieving an event at a specific offset, and (2) \textit{Append}, appending an event to the log. Unlike VMS states, event logs are not owned by any single VMS. Instead, they are logically shared among VMSes: exactly one VMS has append permission, while the others only have read permission.

\noindent\textbf{VMS System.} 
A VMS system consists of a set of $VMS$ and a set of event logs.
For example, a system containing \textit{Cart} and \textit{Product}, managing the state of user carts and product information, respectively, can be modeled as two distinct VMSes, as shown in Fig.~\ref{fig:architecture}. \textit{Cart} replicates some product information managed by \textit{Product}, a recurring pattern in EDMA to reduce latency~\cite{laigner2024benchmark}. \textit{Product} appends product price updates to the event log identified as \textcolor{mauve}{\texttt{update-price}}, prescribed by the mapping $MO$. This makes the update available for \textit{Cart}, and upon reception, the function prescribed by \textit{Cart}'s mapping $MI$ is triggered, updating the product replica.

\fakesub{Key Insight} The VMS model specifies the structures of an EDMA application, components' behaviors, the state and event dependencies, and precise system guarantees. This not only provides developers with tools to define and reason about the architecture and behaviors of their applications, but also enables the underlying system to identify potential bugs during compilation, such as incorrectly specified VMS mappings or updates to replicas of data owned by other components. In contrast, existing popular platforms, such as Dapr and other composite platforms, impose little semantics on the application architecture and hence are unable to support these.



\subsection{Specifying Virtual Micro Services}
\label{subsec:programming}





In this section, we present programming constructs that allow developers to specify a VMS system as well as the sought-after data management requirements. We address the challenges of realizing the VMS model over popular programming abstractions used in practice, lowering its adoption barrier. 
We employ annotations prevalent in SDKs of existing frameworks such as Dapr and Orleans~\cite{orleans_streams}, to specify the input/output mapping of events to functions. VMSes are modeled based on the service~\cite{service} and repository~\cite{repository} patterns, popular in ORM-based applications~\cite{adhoctransactions}. The data access API extends the Java Persistence API (JPA)~\cite{jpa} with familiar concurrency constructs and relieve developers from managing connections, transactions, and locks required by JPA.





\fakesub{Building Blocks}
Listing 1 exhibits an excerpt of the Cart component. Tables are specified by annotating plain objects with the \texttt{@VmsTable} annotation (lines 1-10). Table columns are derived from attributes annotated with \texttt{@Column} or \texttt{@Id}. Both declare an attribute as a column in the table, while the latter indicates that the column is a part of the primary key. Secondary indexes can be declared on arbitrary columns. Lines \ref{line:sid} and \ref{line:pid} declare a composite secondary index on \texttt{seller\_id} and \texttt{product\_id}. Primary and secondary indexes are created automatically using the order of columns. Users can also declare foreign keys on other VMS tables, e.g., line~\ref{line:foreignkey}, and are automatically indexed by \name. Optionally, users can declare constraints on columns, such as line~\ref{line:nonnegative_voucher}, on which a voucher assigned to an item cannot be negative.



An instance of a class annotated with \texttt{@VmsTable}, such as \texttt{CartItem}, abstracts one row through an update operation, as seen in line \ref{line:update}.
Common database operations are abstracted by the \texttt{IRepository} interface (Line \ref{line:repository}), which includes traditional operations such as \texttt{lookupByKey}, \texttt{insert}, \texttt{delete}, and \texttt{update}. 
Users can optionally extend this interface to provide additional predeclared queries, such as line \ref{line:query}. Queries can also be declared in the method body via a query-building API. \name\ can transparently select an available index via an optimizer. For example, in this case, the \textcolor{mauve}{product\_idx} index (declared in lines \ref{line:sid} and \ref{line:pid}) is selected by the \name\ to execute \texttt{getProductByKey} (line \ref{line:byKey}).

\lstset{style=myJava}
\begin{lstlisting}[escapechar=*,caption={Excerpt of Cart component as a VMS},captionpos=b]
|@VmsTable|(name="cart_items") public class CartItem {
   |@Id| |@VmsIndex|(name = "prod_idx")*\label{line:sid}*
   public int seller_id;
   |@Id| |@VmsIndex|(name = "prod_idx")*\label{line:pid}*
   public int prod_id;
   |@Id| |@VmsForeignKey|(table = "customer") *\label{line:foreignkey}*
   public int customer_id;
   |@Column| |@PositiveOrZero|  *\label{line:nonnegative_voucher}*
   public float voucher; // ...
}
interface IProdDB extends IRepository<ProductReplica> { }
interface ICartDB extends IRepository<CartItem> { *\label{line:repository}*
   |@Query|("select *$\ast$* from cart_items 
             where seller_id = :sellerId and prod_id = :prodId") *\label{line:query}*
   List<CartItem> getItems(int sellerId, int prodId); 
}
|@Microservice|("cart") *\label{line:microservice}* public class CartVMS {
   |@Inbound|("price-updated") |@Transactional|(type = RW)*\label{line:tx}**\label{line:inbound}*
   public void processPriceUpdate(PriceUpdated evt) {
      var prod = prodDB.lookupByKey(evt.sellerId, evt.prodId);*\label{line:byKey}*
      prod.price = evt.price; prodDB.update(prod); *\label{line:update}*
      var items = cartDB.getItems(evt.sellerId, evt.prodId);
      for (var cartItem : items) {
         cartItem.voucher += (evt.price - cartItem.unit_price);
         cartItem.unit_price = evt.price; 
      }
      cartDB.updateAll(items);
   } // ...
\end{lstlisting}

A VMS is declared by annotating a class with \texttt{@Microservice} (line \ref{line:microservice}), and event-driven functions of a VMS are declared by the annotation \texttt{@Inbound}. 
The \texttt{@Inbound} annotation and the function parameters reflect which event log triggers the function and the corresponding event type. Users must provide a name to facilitate event log identification, as observed in \texttt{updateProductPrice} (line \ref{line:inbound}). There should be one \texttt{IRepository} interface per VMS table. Instances of \texttt{IRepository} interfaces are created dynamically and made available to functions transparently via the dependency injection technique~\cite{injection}. Thus, users can write arbitrary application logic within a VMS function, as well as query and update the application state using \texttt{IRepository}. Declaring output events is optional, and its absence indicates no events are generated by the VMS function.

A VMS function can be annotated with \texttt{@Transactional} (line \ref{line:tx}) to enable transactional support. Optionally, users can specify whether this function operates in the VMS state in read-write (RW) or read-only (R) mode, allowing for execution optimizations. Unspecified application functions are treated as RW. As seen in \texttt{updateProductPrice}, no begin-transaction or commit statements are necessary. In the case of a constraint violation or program exception, the transaction is aborted without user intervention. \name\ transparently manages transactional concerns, allowing for clean application logic specification. Therefore, developers are not burdened with exception handling or special callbacks.

\begin{lstlisting}[escapechar=*,caption={Excerpt of Product component as a VMS},captionpos=b]
|@Event| class PriceUpdate(int sellerId, int prodId, ...) {
   (int,int) getId(){ return new (sellerId, prodId); }
} *\label{line:partition_key}*
|@Microservice|("product") public class ProductVMS {
   |@Inbound|("update-price") |@Outbound|("price-updated") *\label{line:inbound_2}*
   |@Transactional|(type = RW) |@PartitionBy|(method = "getId") *\label{line:partition}*
   public PriceUpdated updateProductPrice(PriceUpdate evt){ *\label{line:trigger}*
       var prod = prodDB.lookupByKey(evt.seller_id,evt.prod_id);
       prod.price = evt.price; prod.dt = evt.dt;
       prodDB.update(prod);
       return new PriceUpdated(evt.sellerId, evt.prodId, ...*\label{line:output}*);
   } //...
\end{lstlisting}


\fakesub{Composing Multiple VMSes}
The output event log of a VMS function can be the input event log of another VMS, composing the functionalities supported by multiple VMSes ($\S$\ref{subsec:vms}). We illustrate this via the \textit{Product} component exhibited in Listing 2. To start, not all events in a VMS system are necessarily produced by VMSes. Some user-facing events can be specified to trigger a VMS function in response to a user request. For example, the input event \textcolor{mauve}{\texttt{update-price}} in line \ref{line:inbound_2} corresponds to a user request to update the price of a product, triggering the function \texttt{updateProductPrice} in line \ref{line:trigger}.
User-facing events enjoy the same log mechanisms as internally generated events.
The \texttt{updateProductPrice} function generates the event \texttt{PriceUpdated} (line \ref{line:output}). As observed in the method declaration, the output event \textcolor{mauve}{\texttt{price-updated}} matches the input event of Cart in Listing 1, line \ref{line:inbound}. 

\name\ can identify event dependencies of VMSes and schedule them transparently, without requiring any implementation beyond the VMSes' logic. In addition, unlike traditional event log systems~\cite{kafka}, which offer iterator-based (often blocking) interfaces for reading and appending to the log, in VMS, the consumption and production of event logs are handled transparently from the user code. As a result, the application code does not need to block while waiting for and publishing events. Furthermore, applications do not need to manage event offsets explicitly, as in existing systems~\cite{kafka}, which are transparently managed by the system to achieve ACID properties, thereby lowering the development burden. 

\fakesub{Transaction Graphs} Microservices often present multi-component interactions that form topologies~\cite{topology_meta,mucache}, which are usually known in advance~\cite{alibaba}. Built on this observation, declaring a multi-VMS transaction involves specifying a transaction graph, where each vertex represents a VMS function and each edge represents an event log. In other words, a multi-VMS transaction consists of a set of operations performed by functions from different VMSes and a set of read/append operations carried out on a set of logs defined in the corresponding transaction graphs. For example, in a transaction definition, \textcolor{mauve}{\texttt{update-price}} represents an edge, and the vertices are \textit{Product} and \textit{Cart} VMSes.



\fakesub{Parallelization Modes}
VMS transactional functions operate in three modes: single-threaded, partitioned, and embarrassingly parallel modes. While the default single-threaded mode is designed for transactions that require RW access to the entire VMS state, the partitioned and parallel modes allow for concurrent access to the VMS state. In partitioned mode, RW functions can exclusively access specific partitions based on keys, while in parallel mode, R functions or RW functions performing only insert operations can access any data concurrently. 
Since functions that need RW access to the entire state (i.e. key-based access cannot be defined) are often exceptions rather than the norm, 
we anticipate that developers will utilize the different execution modes to optimize the concurrency of VMSes in a similar line as observed in practice~\cite{adhoctransactions,cheng2023developer}.

The VMS model prescribes partitioned and embarrassingly parallel executions of each VMS function through the \texttt{@PartitionBy} and \texttt{@Parallel} annotations, respectively. 
In Listing 2, line \ref{line:trigger}, the method utilized to determine the key-based access is "\textcolor{mauve}{getId}," found in the input event class \texttt{PriceUpdate} (line \ref{line:partition_key}). This means that all data accesses performed by this function will relate to the particular product ID. Annotating a function with \texttt{@Parallel} indicates that the function can run concurrently on multiple input events. More precisely, events triggering partitioned functions accessing different partitions using the same partition method, or those triggering embarrassingly parallel functions, could be processed concurrently. Note that events triggering functions with different modes or partitioned mode using different partition methods would never be processed concurrently. An incorrect specification can create conflicts, which will be detected, and the conflicting transactions will be aborted to achieve safety. Details are discussed in $\S$~\ref{subsec:tx_mgmt}.

\vspace{-1ex}
\subsection{Discussions} 

{\bf Key insights.} A primary issue with current EDA practice is that the application is a black box, not exposing semantic information to the database system to enforce data consistency and integrity. \name\ addresses this challenge by opening up the black box and unveiling the semantic information of transactions and data operations via the VMS model. The VMS model enables developers to define data constraints and transaction dependencies at the application level. \name\ can enforce transactional properties and data constraints transparently and globally, thanks to its cross-stack design that spans the application, messaging, and database layers.

\textbf{The actor model} also offers asynchronous communication and local state encapsulation. However, concurrency in actor systems is achieved through the dynamic creation of actors and pipelining~\cite{agha_actors_1986}, explicitly expressed via application logic. Due to the single-thread execution mode, developers must partition their applications into fine-grained actors. Determining the partitioning scheme, actor placement strategy across nodes, and the number of active actor instances imposes additional application design complexities, often involving performance and data consistency trade-offs, an unsolved research problem~\cite{WangRBSMZ19}. Last but not least, parallelizing a component on multiple actors would also make querying the component state consistently more complicated. State-of-the-art actor systems do not offer querying features across multiple actors. FaaS runtimes follow similar principles, so similar reasoning applies to them. In contrast, \name\ does not require explicit state partitioning of a component during application development. 
Concurrency in \name\ is achieved by partitioning annotations, obviating the cost and complexity of creating and placing new actor instances, and allowing developers to use SQL to query the entire state of a component. 

%% file: sections/04_system.tex
\vspace{-1ex}
\begin{figure}
\centering
\includesvg[width=\columnwidth]{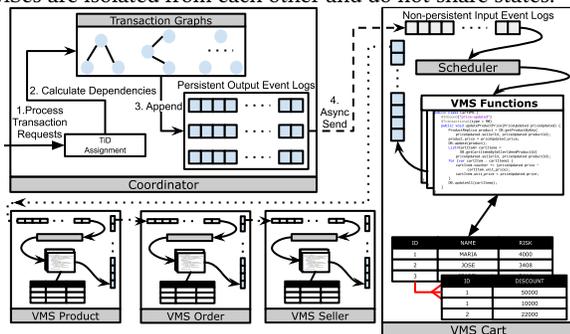}
\vspace{-4ex}
\caption{Architectural Overview of \name}
\label{fig:architecture}
\vspace{-3ex}
\end{figure}

\section{\name\ Architecture}
\label{sec:architecture}

This section presents the architecture of \name, 
a distributed framework that allows developers to design a comprehensive data management solution for EDMA using the VMS model. Its design objective is to achieve transactional data management with high performance while respecting the core EDMA principles. Therefore, it is fundamental for \name\ to minimize inter-VMS coordination, optimize event log processing, and maximize intra-VMS concurrency control, allowing for flexible management and allocation of resources to the VMSes in a manner similar to microservices implementation and deployment in practice.

\name\ is designed and implemented from scratch atop Java Virtual Machine (JVM) runtime using JDK 21 with 35K lines of code. 
We choose JVM because it is a popular runtime for ORM-based microservice applications and offers rich extensibility and customization support to meet the VMS programming constructs ($\S~\ref{subsec:programming}$).

As shown in Figure~\ref{fig:architecture}, the \name\ architecture is organized as a collection of components, namely, VMS instance, proxy, dispatcher, commit handler, and catalogs. Each component is an operating system (OS) process that employs different threads for varied tasks ($\S~\ref{sec:implementation}$), thus they can be independently deployed.
The modular design enables great flexibility in the cloud, allowing users to change deployments without requiring modifications to the application code.
\name\ components exchange events and system messages (e.g., transaction commit and abort) via the network or inter-process communication protocols, according to the deployment model. This is transparent to application code.


\fakesubb{Catalogs} maintain the metadata of VMSes, events, data models, data and event dependencies, and transaction definitions. {\bf Proxies} handle client requests for transaction registration and invocation, check their correctness, and forward them to a dispatcher. {\bf Dispatchers} assign ordered transaction identifiers to multi-VMS transactions, group them into batches, compute the dependency of transactions within each batch, and dispatch transaction invocation events to the corresponding event logs. 
{\bf Commit Handlers} commit or abort transaction batches,
operating independently from dispatchers.

A VMS system in \name\  consists of a collection of standalone \fakesubb{VMS instances}, each mapped to a microservice in a traditional deployment setting. A VMS instance follows a modular design internally with specific-purpose components. The \textbf{transaction scheduler} of a VMS instance relies on the ordered schedule generated by the dispatchers to determine which VMS functions to execute, leveraging the concurrency specifications of the VMS functions to maximize concurrency. This design embodies Inversion of Control (IoC)~\cite{ioc}, in which the framework runtime, rather than the application code, controls how and when VMS functions must execute, the key to achieving serializability of multi-VMS transactions.

The {\bf state management} module manages the relational tables, executes data operations, and enforces data constraints. To maximize concurrent execution of events, it maintains multiple versions of the state updated by concurrent writers while allowing multiple concurrent readers to access consistent snapshots and respecting the transaction isolation property. The {\bf event log management} and state management modules coherently persist the produced events and updated state, respectively, upon the commit of a transaction batch. This design unifies event log management, data management, and event-driven execution into a common framework to achieve efficient ACID transactions across multiple VMSes.



%% file: sections/05_protocols.tex
\vspace{-1ex}
\section{\name\ Protocols}
\label{sec:protocols}

Achieving ACID transactions across components in an EDMA is challenging due to asynchrony and data decentralization. Existing implementations of distributed commit protocols, such as 2-Phase Commit, are not only costly but also exhibit blocking behavior, requiring waits among participating components and typically use communication outside of event channels, jeopardizing asynchrony and decoupling.


To solve this conundrum, we propose an efficient deterministic scheme that maintains the asynchronous message communication abstraction and prevents breaking the sought-after state encapsulation principle of EDMAs. It is based on the key observation that the topology of multi-component transactions of an EDMA application is deterministic and known in advance~\cite{topology_meta,mucache}. However, it does not assume that the VMS functions are deterministic, contrasting the assumptions in deterministic databases~\cite{calvin}.


\vspace{-1ex}
\subsection{Transaction Management}
\label{subsec:tx_mgmt}


\name\ divides transaction scheduling into two parts to maximize performance. The dispatchers reason about transaction dependencies at the VMS level to determine the execution order across VMSes. Next, individual VMSes exploit the semantic information from user specifications to maximize concurrent accesses to data items without conflicting with the transaction order defined by the dispatchers. Different from state-of-the-art deterministic protocols~\cite{calvin}, the read-write set of transactions is not required at any point in \name.



\fakesub{Dispatcher Transaction Scheduling} 
Each transaction request is processed by a dispatcher, responsible for assigning it a unique and ordered TiD and computing its precedent TiD on each VMS it involves. The latter is needed because not every transaction involves all VMS instances.

To allow for scalable transaction dispatching while ensuring the monotonically increasing property of TiD assignment, we divide the TiD namespace into non-intersecting, fixed-size ranges. Inspired by Snapper~\cite{snapper}, dispatchers are placed in a ring, each operating over an exclusive range. TiDs are assigned in the context of a batch. A batch is sealed when a dispatcher runs out of TiDs or a parameterized epoch elapses.
Then, the dispatcher sends the last assigned TiDs of each VMS to the next dispatcher in the ring, which then merges what it receives with its local information and proceeds in the same way.
Upon receiving the information from its predecessor in the ring, a dispatcher has sufficient information to reason about the precedent TiD of each VMS in the batch it is forming. So it can emit the transactions and start processing the next batch.
Dispatchers are given a batch ID to start, and when a batch is sealed, they compute the next batch ID as the current batch ID plus the number of dispatchers.

\fakesub{VMS Transaction Scheduling}
For events involved in multi-VMS transactions, their processing order is scheduled according to the $TiD$s. To eliminate unnecessary coordination, two types of program specifications are leveraged by \name\ to minimize coordination: (1) A VMS function can be declared as read-write (RW) or read-only (R). Within each VMS instance, multiple versions of each data tuple are maintained. Given the initial VMS state $S_0$, after an RW function processes a transaction with $TiD=x$, the state is updated to a valid state $S_x$. 
RW functions never need to wait for the processing of an event triggering an R function with a smaller $TiD$. This is because an event triggering an R function with $TiD=y$ can read the state version $S_z$, the latest version with $z<y$.
(2) The concurrency constructs in functions can further optimize \name\ transactions by allowing for non-conflicting transactions to modify the database state concurrently without coordination. 

\SetKwRepeat{Do}{do}{while}

\begin{algorithm}[tb]
\caption{VMS transaction scheduling}
\label{algo:tx_sched}
\While{true}{
    $t \gets$ getNextBlocking()\;
\label{algoline:getnextblocking}
    \If{t.mode=SINGLE}{\label{algoline:singlethread_begin}
        \If{$\neg(partSched\neq\emptyset \lor paraSched\neq\emptyset) \lor (\exists t:t\in partSched \cup paraSched \land t.type=RW)$}{\label{algoline:singlethread_ok}
            \textbf{block while} above cond. holds\; \label{algoline:singlethread_block}
        }
        \lIf{t.type=R}{\textbf{dispatch}$(t)$;
        \textbf{else} {{\bf execute}$(t)$}}   \label{algoline:singlethread_end} 
        \textbf{continue}\;
    }
    \Do{$t \ne null \land modeAux = t.Mode$}{
        \uIf{$t.mode=PARA \land (partSched=\emptyset \lor (\forall t:t\in partSched \land t.type=R))$}{ \label{algoline:para_begin}
            $paraSched.add(t)$\; 
            \textbf{dispatch}$(t)$\;
        } \label{algoline:para_end}
        \uElseIf{$t.mode=PART \land (paraSched=\emptyset \lor(\forall t:t\in paraSched \land t.type=R))$}{\label{algoline:part_begin} 
        \uIf{$partSched[t.part] = \emptyset \lor (\forall t: t\in partSched[t.part]\land t.type=R$)}{\label{algoline:part_ok}
             $partSched[t.part].add(t)$\;
             \textbf{dispatch}$(t)$\; }\label{algoline:part_dispatch}
        \lElse{ \textbf{block until} $partSched[t.part] = \emptyset$\label{algoline:part_block}}
        }\label{algoline:part_end}
        \lElse{ \textbf{break}}
        $modeAux \gets t.mode$; $t \gets$ getNext()\;
    }
}
\end{algorithm}
Algorithm~\ref{algo:tx_sched} presents the algorithm used by the thread responsible for scheduling transaction events. It retrieves the next transaction from the ordered list based on the ID of the last executed transaction (line~\ref{algoline:getnextblocking}), which is blocked if there is no new transaction to be scheduled. If the transaction event triggers a single-threaded function, it will execute the transaction, or if it is an R function, dispatch it for concurrent execution by a thread pool (lines~\ref{algoline:singlethread_begin}--\ref{algoline:singlethread_end}). The latter can be done because the multi-version database allows a subsequent function to execute concurrently with this R function, as explained above. Transaction events triggering embarrassingly parallel functions (lines~\ref{algoline:para_begin}--\ref{algoline:para_end}) are dispatched concurrently to a thread pool. In addition, events triggering partitioned functions (line~\ref{algoline:part_begin}) accessing different partitions, or accessing the same partition as the previously dispatched events, but all of the latter are R functions (line~\ref{algoline:part_ok}), can be dispatched to a thread pool (line~\ref{algoline:part_dispatch}). Furthermore, it disallows transaction events triggering functions with different modes to execute concurrently unless all the existing transactions are read-only (lines~\ref{algoline:singlethread_ok}, \ref{algoline:para_begin}, and \ref{algoline:part_begin}). The thread blocks if no events can be scheduled (lines~\ref{algoline:singlethread_block} and \ref{algoline:part_block}) to release resources for the other threads to carry out other tasks, e.g., processing events, query execution, and checkpointing ($\S~\ref{sec:implementation}$). 
More details of the algorithm are presented in Appendix ($\S$~\ref{subsec:vms_tx_sched}).

\fakesub{Correctness} We have the following theorem regarding the schedule produced by the VMS. 
The detailed proof is presented in Appendix ($\S$~\ref{subsec:vms_tx_sched}).

\begin{theorem}
Let \( O = [T_1, T_2, \dots, T_n] \) be a total order of transactions defined by the dispatchers, and \( S \) be the schedule produced by the VMS. \( S \) is conflict-equivalent to \( O \).
\end{theorem}

\fakesub{Batch Commit}
Committing a distributed transaction introduces substantial network round-trip times~\cite{consensus}. To mitigate this overhead, transactions commit in batches in \name. 
When a batch is sealed, the dispatcher collects all VMSes that are the sink nodes of the transactions in the batch, logs them in the Catalogs, and sends a $batch\_commit\_info$ to each of them, which includes the IDs of this and the previous batch, and the number of respective TiDs, so every VMS instance can maintain the number of TiDs executed per batch. 

Upon receiving a $batch\_commit\_info$ message, the sink VMSes in the batch wait until they have processed all TiDs. Then, they send a $batch\_complete$ message to the commit handler. When all $batch\_complete$ messages arrive, the commit handler sends a $batch\_commit$ request to all VMSes involved in the batch.
Note that a completed batch in sink VMSes implies that it has also completed in upstream VMSes. This allows for minimizing messages in the batch commit protocol. Besides, transactions are optimistically processed, i.e., VMSes do not wait for batch commit completion in order to process transactions of subsequent batches. 



\fakesub{Transaction Abort}
If a VMS function fails to comply with a database constraint or an incorrectly specified execution mode leads to a conflict, the VMS starts the abort process. It (i) stops scheduling new transactions, (ii) removes the aborted transaction from its set of transactions, (iii) cleans up the write sets of the transactions with TiDs between the aborted TiD and the last executed TiD, and informs the commit handler about the abort. The commit handler logs the abort information (TiD, batch ID) in the Catalogs and multicasts it to VMSes involved in the corresponding transaction (only upstream VMSes are necessary). Upon receiving an abort message, VMSes proceed similarly to the VMS that started the abort procedure, executing steps (i) to (iii) above. 
VMSes must ignore incoming input events that have TiD > aborted TiD until its producer VMS restores the precedence of input events.
Transaction scheduling is resumed from the successor of the aborted transaction in the log. 
Note that at this point, VMS states have not been checkpointed yet, and the event logs written as part of the current batch execution have not been logged yet. The protocol is largely decentralized. It is only necessary to involve the commit handler because upstream VMSes may be unknown to the VMS that initiates the abort procedure. More details of the algorithm is presented in the Appendix ($\S$~\ref{subsubsec:detailed_tx_sched}).



\vspace{-1ex}
\subsection{Durability}

\textbf{Logging}. While SQL statements act as the logical logging constructs for traditional DBMSs, event logs are leveraged as the logical logging in \name. Event logs trigger computations in application components and, in consequence, updates to their internal states. Thus, it is advantageous to leverage the event log rather than the possibly many individual updates to VMS states, allowing for minimizing I/O overhead. 
The logs dispatched to consumer VMSes are placed into a pending logging buffer. Unlike using an external traditional event log system designs~\cite{kafka}, \name\ only needs to persist logs at commit time. The pending logging buffers are flushed upon a batch commit to amortize disk I/O overhead. 






\noindent\textbf{Checkpointing.} Logical logging alone can lead to a prohibitive cost on state recovery because all event logs require replaying to align the VMS states to reflect the last committed batch. Therefore, \name\ features a checkpointing mechanism that is performed at batch commit time. A background thread in each VMS identifies which data item versions are part of the batch pending commit and flushes them all to disk atomically. The process can optionally perform garbage collection, removing old versions that no new transaction will observe in order to reduce memory pressure. By counting on the multi-version VMS database, the checkpoint process does not conflict with any concurrent function or require synchronization with other modules. Due to space constraints, fault-tolerance protocols are discussed in the Appendix ($\S$~\ref{subsec:fault_tolerance}).






\vspace{-1ex}
\subsection{Discussions}


Based on the CAP theorem, \name\ is not an available system for transactions that involve unavailable services. However, the coordinator can mitigate the impact of unavailable services by aborting the impacted transactions and avoiding scheduling new transactions involving the impacted service(s). For transactions that only involve non-impacted VMSes, the \name\ system is still available. If the coordinator is unavailable, multi-VMS transactions cannot be scheduled, but all single-VMS transactions that are read-only or do not conflict with uncommitted batches can be executed. Workflows without transactional annotations behave as in a BASE system (without transactional guarantees), enriching the design choices as developers can make a trade-off depending on applications' requirements.

%% file: sections/06_implementation.tex
\section{Implementation and Optimizations}
\label{sec:implementation}



\begin{figure}[tb]
\centering
\includesvg[width=\columnwidth]{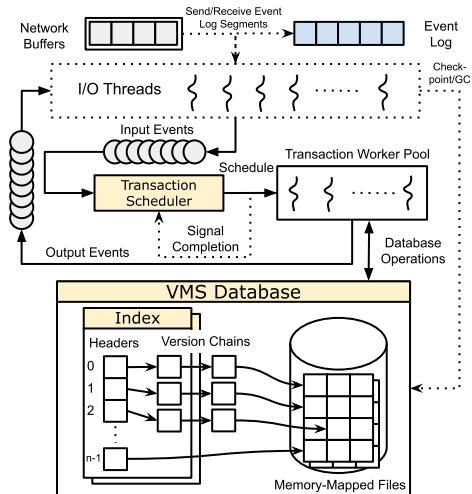}
\vspace{-4ex}
\caption{VMS Internals}
\label{fig:vms}
\vspace{-2ex}
\end{figure}

To provide a high-performance framework that unifies event and data management, numerous optimizations are required, including network I/O, transaction scheduling, concurrent data access, query processing, and storage formats for both data and event logs. Figure~\ref{fig:vms} illustrates the VMS internals we discuss next.

\subsection{Network I/O Processing}
\label{subsec:impl_io}

Network communication with the dispatcher and among VMSes is achieved with TCP connections through asynchronous I/O~\cite{nio}. This applies to both event logs and protocol messages. I/O threads are divided into two categories, receivers and senders. The OS kernel signals receivers about new network I/O events, which in turn 
unmarshal log entries from the buffers, deserialize individual input events, and queue them for transaction processing. Whenever output events are made available, senders dispatch them to the corresponding consumer VMSes. If there is a failure in sending event logs (e.g., VMS crash or network partition), they are put in a pending buffer for later reattempt.

A configurable number of receivers are placed in a network thread pool, returning whenever network events are not available. Senders 
run independently and are uniquely assigned a VMS consumer to send log entries to. This prevents cache pollution, i.e., cache lines of network events evicting cache lines of VMS consumer, and vice versa. In this case, whenever output event logs are not available, senders yield their CPU time to other system threads.

\vspace{-1ex}
\subsection{Transaction Management}

\name\ employs Inversion of Control (IoC), which makes event processing and function scheduling transparent to users as in other distributed frameworks, while enforcing transaction isolation. A transaction scheduler thread runs Algo~\ref{algo:tx_sched} for scheduling VMS functions. It reasons about the concurrency modes specified in the VMS application code ($\S$~\ref{subsec:tx_mgmt}) to maximize throughput while coping with transaction isolation. The transaction scheduler maintains a work-stealing pool of transaction worker threads responsible for executing VMS functions. The work-stealing pool enhances performance because once multiple functions are scheduled to execute, there are no dependencies between them, and, therefore, transaction worker threads are free to "steal" tasks from each other, preventing additional thread parking and unparking costs.
To maximize resource efficiency, the transaction scheduler thread blocks if no new event input is available for scheduling (Algo~\ref{algo:tx_sched}). Upon terminating a function execution, the transaction worker thread updates the scheduler TiD progress, queues the event output, if present, to sender I/O threads, and proceeds to steal a task from the pool.

\fakesub{Transaction Dispatching}
Network I/O processing in dispatchers follows a similar design to a VMS, but is targeted at maximizing transaction request dispatching while not jeopardizing the TiD assignment process.
To meet this goal, sender I/O threads never block. Instead, in the absence of new event inputs, they yield their CPU time to another thread. This ensures minimized waiting times for VMSes because blocking would lead the I/O thread to park and unpark, a costly process for input generation. Receiver I/O threads function in the same way as in a VMS.
Dispatchers assign TiDs to multi-VMS transactions and compute the dependencies of each VMS involved. They operate as independent threads in an event loop mode (i.e. non-blocking), outside the context of a thread pool.


\vspace{-1ex}
\subsection{VMS State Management}
\label{subsubsec:storage}

\name\ provides a purpose-built database to minimize coordination and maximize VMS concurrency. It takes advantage of the substantial increase in main memory sizes observed in recent years, which are often sufficient to accommodate the entire state of transactional systems. 
The co-design of VMS abstraction and the database enables several performance optimizations as explained below.

\noindent\textit{Natively in-process.} 
An in-process database is instantiated for each VMS instance in \name\  to manage its relational tables. Albeit state-of-the-art databases like Duck-DB~\cite{duckdb} operate in-process, applications typically necessitate costly calls to 
database running outside the application process for data access (e.g., via JNI~\cite{jni}). In \name, all components are natively in-process, meaning calls never leave the application's process context. This allows for data transfers between the VMS database and application code with minimal overhead. In addition, data access technologies like ORMs through JDBC~\cite{jdbc} must parse application objects into a database message protocol and vice versa.
\name, however, operates natively with entities, speeding up key tasks such as constraint enforcement, query processing, and durable writes.



\noindent\textit{Multi-versioning.} To optimize the execution of concurrent transactions and ad hoc read-only queries, 
VMS database is designed as a multi-version system. 
Unlike traditional multi-version systems that assign transaction identifiers or transaction timestamps in-flight~\cite{mvcc}, the VMS database relies on the ordered TiDs assigned by the dispatchers to assign versions to data items, ensuring that each transaction appears to be executing alone on a dedicated system.


The use of a specialized multi-version system for a VMS is beneficial for several reasons:
(i) Read-only transactions that do not generate output event logs do not need to be scheduled using the same (thread/CPU) resources allocated to read-write transactions; besides, since their execution does not jeopardize the system progress, their execution can be scheduled at any time as long as the respective versions are available to be read, increasing scheduling flexibilities of the system.
(ii) Even if a read-only transaction generates an event output log, the multi-versioning allows read-only transactions to always be scheduled as embarrassingly parallel, maximizing performance. This would not be possible in a lock-based mechanism.
(iii) Apart from multi-VMS transactions, users may issue ad hoc queries to a VMS instance. The multi-version database enables users to query a consistent snapshot of the VMS state at any moment.

\noindent\textit{Lock-free data structures.} VMS database employs lock-free in-memory data structures for primary and secondary indexes, allowing transaction worker threads to read and create new versions of data items simultaneously without coordination.
Tables are currently implemented as a collection of concurrent hash tables -- single-value hash tables for unique indexes (e.g., primary key indexes) and zero or more multi-value hash tables for secondary indexes. \name\ supports range queries that iterate over primary or secondary index entries. Extending the interfaces to tree structures~\cite{volcano} is left for future work.

The different versions of data items are kept in primary key indexes, and secondary indexes only maintain a pointer to the associated primary index entry. Data item versions are chained as a linked list in a primary key index entry. The head represents the latest data item version. Each node in the list contains the associated TiD representing its version. The versions in the linked list are naturally ordered by \name's concurrency control mechanisms.

\noindent\textit{Query processing.} To take further advantage of the native in-process design and decrease pressure on memory caches, \name\ packages with several predefined query execution plans that employ operator inlining to optimize query execution. If a query does not have a predefined query plan, \name\ can switch to the traditional volcano strategy~\cite{volcano}. In addition, different from ORM libraries that are oblivious to query plans, \name\ integrates both query execution and entity construction into a common execution plan. As a multi-version system, entities are naturally cached and reused to reduce the costs of serializing object results.

\noindent\textit{Data persistence and version deprecation.} A VMS database operates in two modes: in-memory and checkpointing. The former only stores data in memory, while the latter uses a memory-mapped file (MMF) that maps to the last data item versions. \name\ leverages the in-process, multi-versioning, and lock-free design for additional optimizations. As explained later ($\S$~\ref{sec:protocols}), flushing (fresh) and removing (deprecated) data item versions can proceed safely with concurrent transactions and ad hoc queries.

\noindent\textit{Extensibility.} VMS database and storage APIs are modularized to allow for future endeavors in disaggregated storage~\cite{OpenAurora}. While it is possible to design the VMS state management as a thin layer on top of other database systems that expose transactional and relational APIs, and version metadata (for concurrent writers support), this would jeopardize some optimizations discussed above, which corroborates our design choices.


\vspace{-1ex}
\subsection{Event Log Management}
\label{subsec:impl_log}

\name\ applies several optimizations to reduce network and storage I/O overheads that apply to both the dispatcher and VMSes. To decrease the amount of bytes sent over the network, I/O threads serialize the payload of events generated by VMS functions into log entries using a configured serialization format (e.g. compacted JSON).  
To decrease the number of I/Os, the log entries are batched into segments to fit an operating system (OS) page before being sent to the associated VMS consumers. For cache-friendliness, the network buffers are also sized to an OS page. 

Each segment is mapped to a region of memory allocated by the OS (i.e. native memory regions), allowing efficient data transfers between the application and the kernel. These regions of memory are cached and reused across I/O threads. Entries are appended contiguously within the segment and never overflow the page, avoiding page segmentation.

Sender I/O threads employ best-effort batching, i.e., event logs are batched within a segment as much as possible, but the priority is submitting events as fast as possible, so Nagle's algorithm~\cite{nagle} is disabled. The goal is to achieve high throughput without compromising latency, a good trade-off found in the experiments ($\S$~\ref{sec:evaluation}).
Whenever network channels are saturated, the sender I/O threads switch to logging mode and switch back when the network channel is free. This design enables higher resource utilization and efficiency by preventing the creation of additional threads for logging tasks.


\noindent\textbf{Log Segment Design.} A segment of the log is a subset of event logs that follows a compact structure, composed of:
\textit{i}) segment size in bytes (int);
\textit{ii}) number of event logs (int)[1-N];
and \textit{iii}) a sequence of event log entries. An event log has the following format: \textit{i}) message type (byte); \textit{ii}) batch (long), TiD (long), and event identifier (string); \textit{iii}) event payload size (int); \textit{iv}) event payload (byte)[1-N]; \textit{iii}) dependence map size (int); and the \textit{vi}) dependence map payload, the precedence of TiDs for VMSes involved in the transaction. 

As events flow downstream, the VMSes no longer involved in the transaction are removed from the dependence map, allowing for decreased metadata. The offset of an event in the log is the TiD assigned by the dispatcher. Concurrent transactions in VMSes, although never leading to corrupted states, can lead to out-of-order appends to the log. However, this is beneficial because sender I/O threads do not need to coordinate with transaction workers, allowing for higher performance. This is possible because every VMS guarantees that the events logs are processed in TiD order.



\color{black}

%% file: sections/07_experiments.tex
\section{Evaluation}
\label{sec:evaluation}



\subsection{Experimental Settings}

\fakesub{Benchmarks}
\label{subsubsec:benchmarks}
We use two benchmarks throughout the experiments: Online Marketplace (OM)~\cite{laigner2024benchmark} and TPC-C~\cite{tpcc}.
OM is chosen given its ability to stress data management features that EDMAs sacrifices ($\S$~\ref{sec:background}). It models a complex e-commerce application consisting of several EDMA components.
We model a high peak order processing scenario in OM, which stresses the system's performance due to the higher complexity of the \texttt{Customer Checkout} transaction. \texttt{Price Update}, \texttt{Product Delete}, and \texttt{Query Dashboard} correspond to 10\% each and \texttt{Customer} \texttt{checkout} forms 70\% of the transaction ratio.

To demonstrate the versatility of \name, 
we port the standard OLTP benchmark TPC-C~\cite{tpcc} to \name. We partition TPC-C tables into three components: (i) \textit{Warehouse} embodies the customer, district, and warehouse tables; (ii) \textit{Inventory} holds the item and stock tables; and (iii) \textit{Order} holds the order, order item, new order, and history tables. 
Similar to previous work~\cite{snapper,calvin,styx}, we use the \texttt{New Order} and \texttt{Payment} transactions. We decompose the transactions across the components, resembling a real-world partitioned distributed application~\cite{base,vldb2021}. 
Tables in TPC-C are typically partitioned based on the warehouse, which is used for specifying VMS execution modes ($\S$~\ref{app:subsec:concurrency}).
\name\ TPC-C full implementation is found in Appendix ($\S$~\ref{subsec:tpcc_impl}).

\fakesub{Baselines} 
We select Dapr~\cite{dapr} as a baseline system. Dapr ships by default with Redis Streams~\cite{redis_streams} as its event log system~\cite{dapr_redis}. We use default Redis to obtain a baseline through a pure in-memory execution and Redis with durability enabled to log event logs (``EL'' in the plots). Furthermore, we design two different Dapr versions: 
(i) 
Each component is designed as a stateless service, offloading all data management operations to its DBMS. We chose PostgreSQL due to its widespread popularity and support for concurrent writers. Our implementation contains ORM and PostgreSQL optimizations ($\S$~\ref{app:subsec:pg_tuning}) to maximize performance (``Dapr+PG'').
(ii) For a direct comparison with \name, we design a relational in-memory database library for Dapr that mimics PostgreSQL APIs with a design similar to our VMS database ($\S$~\ref{sec:implementation}).

We also use Orleans~\cite{bykov2010orleans} as a baseline, extending the original codebase provided by OM~\cite{laigner2024benchmark} to a distributed setting~\cite{orleans_cluster}. 

\fakesub{Deployment} We use C5n instances of Amazon EC2 deployed in the eu-north-1 region. Unless explicitly noted, each component in Dapr and \name\ is assigned to a specific EC2 instance to prevent resource competition. Orleans is deployed on the same number of EC2 instances as its competitors, but it does not allow for controlling component allocation. Each instance runs Ubuntu 22.04.4 with the kernel 5.10.17 and hyper-threading enabled. The OM benchmark driver, Redis, and dispatchers (together with the proxy, commit handlers, and catalogs) are each deployed in an x9large instance. 

For \name, one dispatcher is used in OM and TPC-C as it is sufficient to sustain the required throughput.
Eight network threads are assigned to the proxy, and a thread is assigned to the commit handler. By default, the batch epoch size is set to 500 ms with a maximum of 10K transactions per batch. 

\fakesub{Methodology} We collect the system throughput and end-to-end latency. The latter is measured as the interval between a transaction being emitted and its commit being received by the client. All experiments are run in six runs of ten seconds each, with the first two as the warm-up runs. After each run, the components' states and event logs are cleaned. To remove cache effects across experiments, we restart the components after each experiment. 

\vspace{-1ex}
\subsection{Transaction Scheduling}
\label{subsubsec:tx_scheduling}

We design a micro benchmark to characterize the performance of \name\ TiD assignment in different benchmarks. This group of experiments was run on a high-end laptop with an Apple M1 chip, 8 cores (4 for performance and 4 for efficiency), and 16Gb RAM. We pre-load the system with enough transactions to prevent TiD assigners from running out of work. For OM, we reuse the transaction ratio specified in $\S$~\ref{subsubsec:benchmarks}. Since the seller dashboard is a read-only query that does not require a TiD, we distribute its percentage uniformly to the other transactions. For TPC-C, we use \texttt{New Order}.

\begin{figure}[tbh]
\centering
  \input{plots/00a_coordinator}
  \input{plots/00b_coordinator}
  \vspace{-3ex}
\caption{Throughput of TiD Assignment}
\label{fig:max_batch_size}
\vspace{-3ex}
\end{figure}
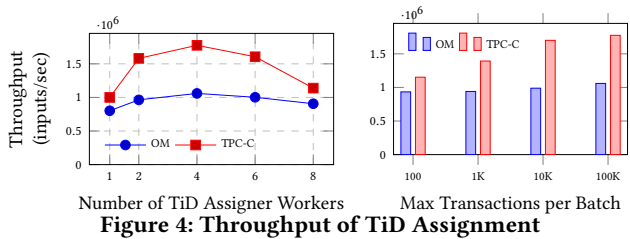

In the first experiment, we fix the batch epoch to infinity and the maximum number of transactions per batch to 100k. Figure~\ref{fig:max_batch_size} (left) shows that increasing the number of dispatchers can provide increased throughput. TPC-C shows higher scalability due to the reduced number of VMSes, resulting in lower overhead on computing dependencies across workers. In any case, the wait time incurred by waiting for dependencies on the ring increases above four workers.

We also verify whether the maximum number of transactions per batch impacts TiD generation performance since the batch size can potentially hide the wait time introduced by the "ring". In this experiment, we fix the number of dispatchers to 4 and vary the batch size. We observe in Figure~\ref{fig:max_batch_size} (right) that the batch size indeed have an impact. For OM, a batch size of 100K improves throughput by 8,8\% compared to 100. Higher improvements are observed in TPC-C.

The experiments highlight the efficiency of the dispatchers design. A single dispatcher generates workloads (with nearly 1 million requests/second), which can stress state-of-the-art OLTP DBMSs and is sufficient for a \name\ system. As a result, the dispatchers cannot be the bottleneck for a useful \name\ setup.

\subsection{Online Marketplace}

\begin{figure}[tb]
\centering
  \input{plots/01_conc_lvl}
\vspace{-3ex}
\caption{Concurrency Overhead with Fixed Resources}
\Description{The plot shows that \name can sustain higher throughout even when the number of clients increase substantially compared to competitors like Dapr and Orleans, that exhibit much lower throughput when facing increased concurrency overhead}
\label{fig:concurrency}
\vspace{-4ex}
\end{figure}

\fakesub{Effect of Concurrency Level}
The \textit{concurrency level} is the maximum number of current worker threads submitting transactions to the system. Apart from capturing the sensitivity to varied concurrency degrees, this experiment is useful to identify the parameter that yields the optimal performance in each system for subsequent experiments. Thus, we maximize the resources available 
to each OM component, and set the workload skewness to be uniform for both Seller and Product.

As shown in Figure~\ref{fig:concurrency}, \name\ shows increasing throughput as the concurrency level increases. The maximum throughput is achieved at the highest concurrency level (36), but from 28 on, the throughput stabilizes as the system is saturated. 
On the other hand, we observe a substantial performance difference in Dapr. 
In Dapr+PG, the workload is dominated by I/Os between the OM components and PostgreSQL, including inter-process communication, object (de-)serialization, and connection management, exhibiting the lowest throughput. Since PostgreSQL and Npgsql are not a native part of Dapr, we do not include them in further experiments.




With the in-memory execution, the I/O overhead found in Dapr+ PG is dramatically reduced. The remaining sources of overhead in Dapr can be explained by two factors: the sidecar design~\cite{sidecars} and the remote event log system. 
The sidecar is an autonomous process that mediates external messages to and from a Dapr component, resulting in several additional threads in addition to the application threads, leading to inter-process communication and context switch overheads,
not to mention the intensive network I/O of event logs.

Orleans' performance is impacted by two factors. To carry out a \texttt{Price Update}, the product actor must resort to Orleans Streams (OS) to deliver the update to the corresponding cart actors to avoid the prohibitive cost of querying all carts. 
However, OS introduces a $60$\% overhead on average due to additional OS actors and inter-node event streams. 
Besides, since Orleans lacks query functionality~\cite{query_orleans}, \texttt{Seller Dashboard} (SD) is computed using C\# query operators~\cite{linq}, adding an average of $64$\% overhead.
%
The remaining overhead can be attributed to the required coordination among a high number of actors~\cite{orleans_best_practices}~, the actor-to-node assignment policy~\cite{orleans_load_balancing}, and the overhead of identifying actor through the directory~\cite{grain_directory}.

\begin{figure}
\captionsetup[subfigure]{labelformat=empty}
\subfloat[]{
  \input{plots/02_scalability}
}
\subfloat[]{
  \input{plots/05_latency}
}    
\vspace{-6ex}
\caption{Scalability (left) and Breakdown Latency (right)}
\label{fig:scalability}
\vspace{-3ex}
\end{figure}
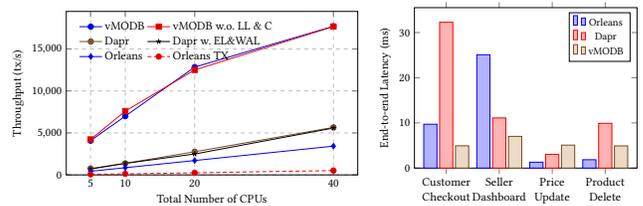

We also collect the latency breakdown of the OM transactions. Figure~\ref{fig:scalability} (right) exhibits the numbers collected with the highest concurrency level. 
Being a long-running and complex transaction, the \texttt{Customer Checkout} latency in Dapr stands out substantially. \texttt{Product Delete} and \texttt{Price Update} transactions involve fewer components and are less prone to increased latency in Dapr and Orleans. Similar trends are found in~\cite{laigner2024benchmark}.
\name\ latencies remain stable due to the batch commit protocol, which we explore later in this section. 

Lastly, \texttt{Seller Dashboard} exhibits lower latency in \name, demonstrating the beneficial aspects of the VMS database ($\S$~\ref{sec:implementation}) over C\# query operators~\cite{linq}.
It is worth noting that Dapr and Orleans provide no transaction isolation for queries, whereas \name\ offers users a consistent snapshot.


Furthermore, the logical logging and checkpointing mechanisms impose little overhead on both \name\ and Dapr. In \name, logical logging (``LL'' in the plots) adds 3,5\% overhead and checkpointing (``C'' in the plots) adds another 3,2\%, demonstrating the effectiveness of optimizing for I/O in \name\ design ($\S$~\ref{sec:implementation}). 
We omit Orleans persistence~\cite{orleans_persistence} results because performance is on par with Dapr+PG.


\fakesub{Scalability}
This experiment compares the scalability of \name, Dapr, Orleans, and Orleans Transactions (TX). We focus on the \texttt{Customer Checkout} because it is the most complex transaction in OM, imposing the highest coordination overhead~\cite{smsa,laigner2024benchmark}. We vary the amount of resources available to each OM component and maximize the concurrency level. 
Figure~\ref{fig:scalability} (left) shows all systems scale as more resources are added, with \name\ being superior in every instance despite ensuring ACID. Besides, \name\ with limited resources is competitive with Dapr and outperforms Orleans when both have full resources available. It is noteworthy that in Dapr and Orleans, application functions operate as embarrassingly parallel, whereas in \name\ the transaction scheduling requires accumulating events in order to comply with transaction isolation, demonstrating the effectiveness of the overhead minimization strategies adopted ($\S$~\ref{sec:implementation}).


\fakesub{Effect of Workload Skewness}
In OM, a highly skewed workload leads to some sellers and associated products being more popular than others. We follow the OM experimental procedure~\cite{laigner2024benchmark}, on which varying seller skewness is sufficient to stress a target system.
Dapr is omitted because it does not support transactional isolation. 

\begin{figure}
\centering
  \input{plots/04c_skewness}
  \input{plots/04d_skewness}
\vspace{-2ex}
\caption{Effect of Skew Levels on \name\ Throughput}
\label{fig:skewness_1}
\vspace{-3ex}
\end{figure}
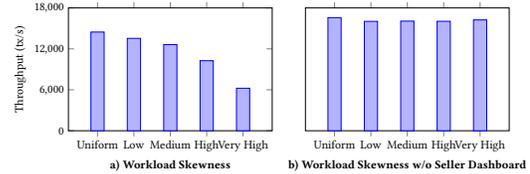

As shown in Figure~\ref{fig:skewness_1}a), the increasing contention leads to a decay in throughput. This can be attributed to the more item versions in the system, increasing memory pressure, and the high tail latency of \texttt{Seller Dashboard}, an avoidable phenomenon given that very few specific sellers and their products are involved in queries. 
In Figure~\ref{fig:skewness_1}b), removing \texttt{Seller Dashboard} from the transaction ratio withdraws the decaying effect, yielding an increase of 18\% (uniform) up to 61\% (very high) in throughput. We also observe that the skewness has a marginal effect on throughput, demonstrating the effectiveness of \name\ concurrency primitives.

\begin{figure}[t]
\centering
  \input{plots/04a_skewness}
  \input{plots/04b_skewness}
\vspace{-2ex}
\caption{Effect of Skew Levels on \name\ Latency}
\label{fig:skewness_2}
\vspace{-2ex}
\end{figure}
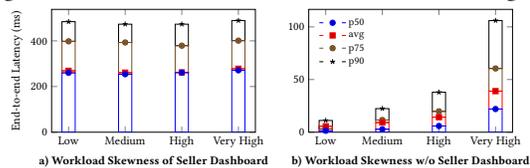

Figure~\ref{fig:skewness_2} exhibits the end-to-end latency for workloads in Figure~\ref{fig:skewness_1}a). We distinguish the results of the \texttt{Seller Dashboard} (\textbf{a}) from the transactions that run in the context of a batch (\textbf{b}) for a proper analysis.
\texttt{Seller Dashboard}'s latency shows little variation across skew levels, whereas that of batch committed transactions increases as skewness increases. 

\subsection{TPC-C}
\label{subsec:tpcc}

We run different mixes of TPC-C transactions to compare \name\ with baselines. The number of warehouses is set to 32 to utilize the available multi-core concurrency of the instance. All the VMS functions are executed conservatively in the single-threaded mode in the first experiment. As the latency results are similar to the previous experiments, we omit them here and put them in the appendix. We compare with two additional baselines: (1) Snapper~\cite{snapper}, a transaction library on Orleans that has demonstrated superior performance over Orleans Transactions, and (2) Styx~\cite{styx}, a state-of-the-art framework for transactional stateful functions, which outperforms the other stateful function frameworks such as Boki~\cite{boki} and Beldi~\cite{beldi}. The implementations of TPC-C on these baselines are described in~\cite{snapper} and~\cite{styx}, respectively. We run \name\ with the same transaction mix and ratio as reported by the baselines. In particular, Styx supports both \texttt{New Order} and \texttt{Payment} transactions, and Snapper only supports \texttt{New Order}. Besides, while Snapper is run with a comparable amount of computational resources to \name, the result of Styx is based on the maximum achievable performance reported in~\cite{styx} with 112 physical CPUs, much more than \name's deployment (38 comparable CPUs).

Figures~\ref{fig:tpcc_throughput}a) and b) show that \name\ outperforms the other two even with the conservative single-threaded mode for all VMS functions. Snapper inherits the overhead of Orleans' virtual actor model, including searching the distributed directory to locate an actor during method invocations, and the overhead of frequent communication between the fine-grained actors. Styx relies on an external event log system to communicate between components, incurring much higher overhead than \name's unified architecture. 

While the previous experiment demonstrated \name\ can achieve competitive performance by only using the single-threaded ({\bf ST}) execution mode, in the next experiment, we study the effect of the other execution modes. We add the following execution mode to the specific VMSes: (1) {\bf PI}:  {\bf P}artitioned mode for {\bf I}nventory based on the warehouse ID, (2) {\bf PO}: {\bf P}arallel mode for {\bf O}rder, and (3) {\bf PW}: {\bf P}artitioned mode for {\bf W}arehouse based on the warehouse ID and district ID. All of them preserve the isolation of transactions. Figure~\ref{fig:tpcc_throughput}c) shows that the partitioned and parallel modes can effectively utilize the available CPU cores to increase the concurrency, thereby achieving much higher throughput than the ST mode.

\begin{figure}
\centering
    \input{plots/10a_tpcc_comparison_styx}
    \input{plots/10b_tpcc_comparison_snapper}
    \input{plots/11_impact_execution_modes}
\vspace{-2ex}
\caption{Throughputs of TPC-C}
\label{fig:tpcc_throughput}
\vspace{-4ex}
\end{figure}

\begin{wrapfigure}{l}{0.455\columnwidth}
\vspace{-2ex}
\centering
    \input{plots/08_scalability_tpcc_new}
\vspace{-5ex}
\caption{TPC-C Scalability}
\label{fig:scalability_tpcc}
\vspace{-2ex}
\end{wrapfigure}

To understand the effect of scale factor and durability in TPC-C, we run \name\ with and without logging and checkpointing enabled.
Figure~\ref{fig:scalability_tpcc} shows that the throughput of TPC-C increases with more warehouses due to the increasing concurrency. Durability imposes, on average, 17\% performance overhead, which reduces with more warehouses (13\% in 16 and 32 warehouses).


%% file: plots/00a_coordinator.tex
\begin{tikzpicture}[scale=0.91]
    \begin{axis}[
        width=5.15cm,height=3.5cm,
        xlabel={Number of Dispatchers},
        xlabel style={font=\small,align=center},
        ylabel={Throughput (inputs/sec)},
        ylabel style={font=\small,text width=1.7cm,align=center},
        tick label style={font=\tiny},
        ymin=500000,
        xtick={1, 2, 4, 6, 8},
        legend style={
            legend columns=2,
            fill=none,
            draw=none,
            legend pos=south west,
            font=\tiny},
        legend cell align={left},
        ymajorgrids=true,
        xmajorgrids=true,
        grid style=dashed
        ]
        

        \addplot
            coordinates {
            (1,801785)(2,964103)(4,1059628)(6,1002110)(8,906128)
            };
        \addplot
            coordinates {
            (1,999990)(2,1581197)(4,1775701)(6,1605369)(8,1138622)
            };
        \legend{OM, TPC-C}
    \end{axis}
\end{tikzpicture}

%% file: plots/00b_coordinator.tex
\begin{tikzpicture}[scale=0.91]
\begin{axis}[
    width=5cm,height=3.5cm,
    ybar=2pt,
    bar width=0.15,
    scaled x ticks = false,
    ylabel style={font=\small,text width=1.7cm,align=center},
    tick label style={font=\tiny},
    xlabel=Max Transactions per Batch,
    xlabel style={font=\small,align=center},
    ymin=0, 
    xtick={1,2,3,4},
    xticklabels={100,1K,10K,100K},
    legend style={
        legend columns=2,
        fill=none,draw=none,
        legend pos=north west,
        font=\tiny
    }
    ]
    \addplot 
	coordinates {
		 (1,933967)(2,940007)(3,988787)(4,1059628)};
    \addplot 
	coordinates {
		 (1,1152406)(2,1393075)(3,1701222)(4,1775701)};
    \legend{OM, TPC-C}
\end{axis}
\end{tikzpicture}

%% file: plots/01_conc_lvl.tex
\begin{tikzpicture}[scale=0.73]
    \begin{axis}[
        width=10.20cm,height=4.5cm,
        scale only axis,
        scaled y ticks = false,
        yticklabel = {
            \pgfmathparse{\tick/1000}
            \pgfmathprintnumber{\pgfmathresult}\,K
        },
        xlabel={Concurrency Level},
        ylabel={Throughput (tx/s)},
        tick label style={font=\tiny},
        ylabel near ticks,
        xlabel near ticks,
        ymin=0, 
        ymax=17000,
        xtick={1, 4, 8, 12, 16, 20, 24, 28, 32, 36},
        ytick={0,5000,10000,15000},
        legend style={
        font=\scriptsize,
        legend columns=4,
        fill=none,
        draw=none,
        at={(0.012,0.85)},anchor=west
        },
        legend cell align={left},
        ymajorgrids=true,
        xmajorgrids=true,
        grid style=dashed
        ]
        \addplot
            coordinates {
            (1,485)(2,922)(4,1600)(8,2800)(12,3900)(16,4800)(20,4800)(24,4900)(32,5096)(36,5091)
            };
        \addplot
            coordinates {
            (1,462)(2,909)(4,1591)(8,2747)(12,3682)(16,4566)(20,5183)(24,5531)(32,5744)(36,6200)
            };
        \addplot
            coordinates {
            (1,469)(2,933)(4,1596)(8,2895)(12,3900)(16,4436)(20,4641)(24,4780)(32,4812)(36,4828)
            };
        \addplot
            coordinates {
            (1,143)(2,275)(4,490)(8,650)(12,866)(16,905)(20,900)(24,902)(28,877)(32,893)(36,854)
            };  
        \addplot
            coordinates {
             (1,1442)(2,2916)(4,5267)(8,8568)(12,10518)(16,12009)(20,12742)(24,13248)(28,13566)
             (32,13496)
             (36,13943)
            };
        \addplot
            coordinates {
             (1,1360)(2,2860)(4,5336)(8,8297)(12,10462)(16,11678)(20,12800)(24,13121)(28,13938)
             (32,13654)
             (36,14026)
            };
        \addplot
            coordinates {
             (1,1396)(2,2910)(4,5255)(8,8442)(12,10328)(16,11808)(20,12704)(24,13285)(28,13683)
             (32,13553)
             (36,13527)
            };
        \addplot
            coordinates {
             (1,84)(2,187)(4,409)(8,803)(12,1094)(16,1375)(20,1496)(24,1627)(28,1680)(32,1702)(36,1648)
            };
        \addplot
            coordinates {
             (1,400)(2,800)(4,1600)(8,2800)(12,3810)(16,4020)(20,4120)(24,4220)(28,4320)(32,4342)(36,4292)
            };
        \legend{Dapr w.o. EL\&WAL, Dapr w. EL, Dapr w. EL\&WAL, Dapr+PG, \name\ w.o. LL\&C, \name\ w.o. C, \name, Orleans, 
        Orleans w.o. OS\&SD}
    \end{axis}
\end{tikzpicture}

%% file: plots/02_scalability.tex
\begin{tikzpicture}[scale=0.5]
    \begin{axis}[
        width=7.75cm,
        height=4.35cm,
        scale only axis,
        scaled y ticks = false,
                yticklabel = {
            \pgfmathparse{\tick/1000}
            \pgfmathprintnumber{\pgfmathresult}\,K
        },
        xlabel={Total Number of CPUs},
        ylabel={Throughput (tx/s)},
        ylabel near ticks,
        xlabel near ticks,
        ymin=0, 
        xtick={5,10,20,40},
        legend style={
        fill=none,
        draw=none,
        legend columns=2,
        at={(0.012,0.85)},anchor=west
        },
        legend cell align={left},
        ymajorgrids=true,
        xmajorgrids=true,
        grid style=dashed
        ]
        \addplot
            coordinates {
            (5,4075) 
            (10,7001) 
            (20,12836) 
            (40,17700) 
            };
        \addplot
            coordinates {
            (5,4225) 
            (10,7596) 
            (20,12501) 
            (40,17668) 
            };
        \addplot
            coordinates {
            (5,755) 
            (10,1375) 
            (20,2756) 
            (40,5660) 
            };
        \addplot
            coordinates {
            (5,694) 
            (10,1374) 
            (20,2489) 
            (40,5581) 
            };
        \addplot
            coordinates {
            (5,428) 
            (10,856)
            (20,1713) 
            (40,3426) 
            };
        \addplot
            coordinates {
            (5,62.5) 
            (10,125)  
            (20,250) 
            (40,518) 
            };
        \legend{
        \name,
        \name\ w.o. LL\&C,
        Dapr, Dapr w. EL\&WAL,
        Orleans,
        Orleans TX
        }
    \end{axis}
\end{tikzpicture}

%% file: plots/05_latency.tex
\begin{tikzpicture}[scale=0.5]
\begin{axis}[
  width=7.5cm,height=5.85cm,
	x tick label style={text width=1.5cm,align=center},
	ylabel=End-to-end Latency (ms),
	enlarge x limits=0.2,
    ymin=0, 
    legend style={
        draw=none,
        legend pos=north east
    },
	ybar,
    xtick={1,2,3,4},
    xticklabels={Customer Checkout, Seller Dash., Price Update, Product Delete},
]
    \addplot 
	coordinates {
        (1,9.71)(2,25.09)(3,1.30)(4,1.85)
        };
    \addplot 
	coordinates {
        (1,32.28)(2,11.10)(3,3.05)(4,9.90)
        };
    \addplot 
    coordinates {
        (1,4.93)(2,7.02)(3,5.08)(4,4.91)
    };
\legend{Orleans,Dapr,\name}
\end{axis}
\end{tikzpicture}

%% file: plots/04c_skewness.tex
\begin{tikzpicture}[scale=0.5]
\begin{axis}[
  width=7cm,height=4.85cm,
    ybar=2pt,
    yticklabel = {
        \pgfmathparse{\tick/1000}
        \pgfmathprintnumber{\pgfmathresult}\,K
    },
    scaled y ticks = false,
    scaled x ticks = false,
	ylabel=Throughput (tx/s),
    xtick={0,1,2,3,4},
    xlabel=\textbf{a) Workload Skewness},
    ymin=0, 
    ymax=18000,
    xticklabels={Uniform,Low,Medium,High,Very High},
    legend style={
        fill=none,
        draw=none,
        legend columns=-1,
        legend pos=north east,font=\small
    }
]
\addplot 
	coordinates {
   (0,14465)
   (1,13539)(2,12623)(3,10267)(4,6228)};
\end{axis}
\end{tikzpicture}

%% file: plots/04d_skewness.tex
\begin{tikzpicture}[scale=0.5]
\begin{axis}[
  width=7cm,height=4.85cm,
    ybar=2pt,
    ymin=0, 
    ymax=18000,
    ytick={0,4000,8000,12000,16000},
    scaled y ticks = false,
    scaled x ticks = false,
    ytick=\empty,
    xtick={0,1,2,3,4},
    xlabel=\textbf{b) Workload Skewness w/o Seller Dashboard},
    xticklabels={Uniform,Low,Medium,High,Very High},
    legend style={
        fill=none,
        draw=none,
        legend columns=-1,
        legend pos=north east,font=\small
    }
]
\addplot 
	coordinates {
		 (0,16571)(1,16028)(2,16065)(3,16040)(4,16243)};
\end{axis}
\end{tikzpicture}

%% file: plots/04a_skewness.tex
\begin{tikzpicture}[scale=0.5]
\begin{axis}[
  width=7cm,height=4.85cm,
        stack plots=y,
        /tikz/ybar,
        ymin=0,
        xlabel=\textbf{a) Workload Skewness of Seller Dashboard},
        ylabel=End-to-end Latency (ms),
        xtick=data,
        xticklabels={Low,Medium,High,Very High},
        legend style={legend columns=4,fill=none,draw=none,
        legend pos=north west
        }
        ]
        \addplot
            coordinates {
            (1,260)(2,254)(3,261)(4,271)
            };
        \addplot
            coordinates {
            (1,9)(2,7)(3,1)(4,7)
            };
        \addplot
            coordinates {
            (1,129)(2,132)(3,117)(4,123)
            }; 
        \addplot
            coordinates {
            (1,86)(2,80)(3,94)(4,88)
            };
    \end{axis}
\end{tikzpicture}

%% file: plots/04b_skewness.tex
\begin{tikzpicture}[scale=0.5]
\begin{axis}[
  width=7cm,height=4.85cm,
        stack plots=y,
        /tikz/ybar,
        ymin=0,
        xlabel={\textbf{b) Workload Skewness w/o Seller Dashboard}},
        xtick=data,
        xticklabels={Low,Medium,High,Very High},
        legend style={legend columns=1,fill=none,draw=none,
        legend pos=north west,font=\normalsize
        }
        ]
        \addplot
            coordinates {
            (1,1.38)(2,2.84)(3,5.9)(4,22)
            };
        \addplot
            coordinates {
            (1,4.27)(2,6.36)(3,8.4)(4,17)
            };
        \addplot
            coordinates {
            (1,-2.5)(2,2.37)(3,5.5)(4,21.5)
            }; 
        \addplot
            coordinates {
            (1,7.98)(2,10.82)(3,18.1)(4,45.5)
            };
        \legend{p50,avg,p75,p90}
    \end{axis}
\end{tikzpicture}

%% file: plots/10a_tpcc_comparison_styx.tex
\begin{tikzpicture}[scale=0.5]
\begin{axis}[
    width=4.5cm,
    height=3.75cm,
    ybar=4pt,
    enlarge x limits=0.5,
    x tick label style={text width=2cm,align=center},
    xlabel=\textbf{a) Comparison with Styx},
	ylabel=Throughput (tx/s),
    xtick={0,1,2},
    ymin=0,
    xticklabels={
        Styx (112 CPUs),
        \name\ (38 CPUs)
    },
    ]
]
\addplot
    coordinates {
        (0,3500)
        (1,25000)
    };
\end{axis}
\end{tikzpicture}

%% file: plots/10b_tpcc_comparison_snapper.tex
\begin{tikzpicture}[scale=0.5]
\begin{axis}[
   width=4.5cm,
  height=3.75cm,
    ybar=4pt,
    enlarge x limits=0.5,
    x tick label style={text width=2cm,align=center},
    xlabel=\textbf{b) Comparison with Snapper},
    xtick={0,1,2},
    ymin=0,
    xticklabels={
        Snapper (40 CPUs),
        \name\ (38 CPUs)
    },
    ]
]
\addplot
    coordinates {
        (0,12000)
        (1,21033)
    };
\end{axis}
\end{tikzpicture}

%% file: plots/11_impact_execution_modes.tex
\begin{tikzpicture}[scale=0.5]
\begin{axis}[
   width=8cm,
  height=4.3cm,
    ybar=4pt,
    xlabel=\textbf{c) Impacts of execution modes},
    xtick={0,1,2,3,4,5,6},
    ymin=0, 
    xticklabels={
        ST,
        PI, 
        PI+PO,
        PI+PO+PW 
    },
]
\addplot
    coordinates {
        (0,21033)(1,31369)(2,60276)
        (3,80000) 
        
    };
\end{axis}
\end{tikzpicture}

%% file: plots/08_scalability_tpcc_new.tex
\begin{tikzpicture}[scale=0.625]
    \begin{axis}[
        width=7.5cm,
        height=4.35cm,
        xlabel={Number of Warehouses},
        ylabel={Throughput (tx/s)},
        xtick=data,
        symbolic x coords={1,2,4,8,16,32},
        legend style={
        legend pos=north west,
        legend columns=1,fill=none,draw=none,
        font=\scriptsize
        },
        legend cell align={left},
        ymajorgrids=true,
        xmajorgrids=true,
        grid style=dashed
    ]
        \addplot
            coordinates {
            (1,25803) 
            (2,30405) 
            (4,47037)
            (8,66428)
            (16,76098) 
            (32,88070) 
            };  
        \addplot
            coordinates {
            (1,21071)
            (2,25479) 
            (4,44114)
            (8,59856)
            (16,69348)
            (32,80000)
            }; 
        \legend{\name\ w.o. LL \& C, \name\ }
    \end{axis}
\end{tikzpicture}

%% file: sections/09_conclusion.tex
\section{Conclusion}

\name\ is a distributed framework that facilitates EDMA application development with high data consistency and integrity. 
Contrasting with the conventional tiered architecture of data-intensive applications~\cite{feral}, 
\name, with its VMS model and unified state and event log management, maintains the core EDMA principles and the appealing simplicity of a stateless program design for developers while achieving transactional data management, strong data consistency properties, and notable performance superior to state-of-the-art solutions.

%% file: sections/99_appendix.tex
\section{Appendix}


\subsection{\name\ TPC-C Implementation}
\label{subsec:tpcc_impl}

\begin{lstlisting}[escapechar=*,caption={VMS Warehouse},captionpos=b]
@Inbound(values = "new-order-ware-in")
@Outbound("new-order-ware-out")
@Transactional(type = RW)
@PartitionBy(clazz = NewOrderWareIn.class, method = "getId")
public NewOrderWareOut processNewOrder(NewOrderWareIn in) {
    District district = this.districtDB.lookupByKey(new District.DistrictId(in.d_id, in.w_id));
    float w_tax = this.warehouseDB.getWarehouseTax(in.w_id);
    district.d_next_o_id++;
    this.districtDB.update(district);
    float c_discount = this.customerDB.getDiscount(in.c_id, in.d_id, in.w_id);
    return new NewOrderWareOut(in.w_id, in.d_id, in.c_id, in.itemsIds, in.supWares, in.qty, in.allLocal, w_tax, district.d_next_o_id, district.d_tax, c_discount);
}
@Inbound(values = "payment-in")
@Outbound("payment-out")
@Transactional(type = RW)
@PartitionBy(clazz = PaymentIn.class, method = "getId")
public PaymentOut processPayment(PaymentIn in) {
    District district = this.districtDB.lookupByKey(new District.DistrictId(in.d_id, in.w_id));
    district.d_ytd += in.amount;
    Warehouse warehouse = this.warehouseDB.lookupByKey(in.w_id);
    warehouse.w_ytd += in.amount;
    this.districtDB.update(district);
    this.warehouseDB.update(warehouse);
    Customer customer;
    if(in.by_name){
        List<Customer> customers = this.customerDB.getCustomerByLastName(in.c_d_id, in.c_w_id, in.c_last);
        int index = customers.size() / 2;
        if (customers.size() % 2 == 0) index -= 1;
        customer = customers.get(index);
    } else {
        customer = this.customerDB.lookupByKey(new Customer.CustomerId(in.c_id, in.d_id, in.w_id));
    }
    if (customer.c_credit.equals("BC")) {
        customer.c_data = "%d %d %d %d %d %f | %s".formatted(customer.c_id, in.c_d_id, in.c_w_id, in.d_id, in.w_id, in.amount, customer.c_data);
        if (customer.c_data.length() > 500) {
            customer.c_data = customer.c_data.substring(0, 500);
        }
    }
    customer.c_balance -= in.amount;
    customer.c_ytd_payment += in.amount;
    customer.c_payment_cnt += 1;
    this.customerDB.update(customer);
    String h_data = "%s    %s".formatted( warehouse.w_name.length() > 10 ? warehouse.w_name.substring(0, 10) : warehouse.w_name, district.d_name.length() > 10 ? district.d_name.substring(0, 10) : district.d_name );
    return new PaymentOut(in.w_id, in.d_id, in.c_id, in.c_w_id, in.c_d_id, in.amount, h_data);
}
@Inbound(values = "order-status-in")
@Outbound("order-status-out")
@Transactional(type = R)
public OrderStatusOut processOrderStatus(OrderStatusIn in) {
    if(in.by_name){
        List<Customer> customers = this.customerDB.getCustomerByLastName(in.d_id, in.w_id, in.c_last);
    } else {
        Customer customer = this.customerDB.lookupByKey(new Customer.CustomerId(in.c_id, in.d_id, in.w_id));
    }
    return new OrderStatusOut(in.w_id, in.d_id, in.c_id);
}
\end{lstlisting}
\vspace{-1ex}

\begin{lstlisting}[escapechar=*,caption={VMS Inventory},captionpos=b]
@Inbound(values = "new-order-ware-out")
@Outbound("new-order-inv-out")
@Transactional(type = RW)
@PartitionBy(clazz = NewOrderWareOut.class, method = "getId")
public NewOrderInvOut processNewOrder(NewOrderWareOut in) {
    float[] prices = this.itemDB.getPricePerItemId(in.itemsIds);
    String[] ol_dist_info = new String[in.itemsIds.length];
    List<Stock> stockItemsToUpdate = new ArrayList<>(prices.length);
    for(int i = 0; i < in.itemsIds.length; i++){
        Stock stock = this.stockDB.lookupByKey(new Stock.StockId(in.itemsIds[i], in.supWares[i]));
        ol_dist_info[i] = stock.getDistInfo(in.d_id);
        int ol_quantity = in.qty[i];
        if(stock.s_quantity > ol_quantity){
            stock.s_quantity = stock.s_quantity - ol_quantity;
        } else {
            stock.s_quantity = stock.s_quantity - ol_quantity + 91;
        }
        stock.s_ytd = stock.s_ytd + ol_quantity;
        stock.s_order_cnt++;
        if(stock.s_w_id != in.w_id){
            stock.s_remote_cnt++;
        }
        stockItemsToUpdate.add(i, stock);
    }
    this.stockDB.updateAll(stockItemsToUpdate);
    return new NewOrderInvOut(in.w_id, in.d_id, in.c_id, in.itemsIds, in.supWares, in.qty, in.allLocal, in.w_tax, in.d_next_o_id, in.d_tax, in.c_discount, prices, ol_dist_info);
}
\end{lstlisting}
\vspace{-1ex}

\begin{lstlisting}[caption={VMS Order},captionpos=b]
@Inbound(values = "new-order-inv-out")
@Transactional(type = W)
@Parallel
public void processNewOrder(NewOrderInvOut in){
    Order order = new Order(in.d_next_o_id, in.d_id, in.w_id, in.c_id, new Date(), in.itemsIds.length, in.allLocal);
    NewOrder newOrder = new NewOrder(in.d_next_o_id, in.d_id, in.w_id);
    this.orderDB.insert(order);
    this.newOrderDB.insert(newOrder);
    List<OrderLine> orderLinesToInsert = new ArrayList<>(in.itemsIds.length);
    for(int i = 0; i < in.itemsIds.length; i++){
        float ol_amount = (in.qty[i] * in.itemsIds[i] * (1 + in.w_tax + in.d_tax) * (1 - in.c_discount));
        OrderLine orderLine = new OrderLine(in.d_next_o_id, in.d_id, in.w_id,i+1, in.itemsIds[i], in.supWares[i],null, in.qty[i], ol_amount, in.ol_dist_info[i]);
        orderLinesToInsert.add(i, orderLine);
    }
    this.orderLineDB.insertAll(orderLinesToInsert);
}

@Inbound(values = "payment-out")
@Transactional(type = W)
@Parallel
public void processPayment(PaymentOut out){
    History history = new History(out.c_id, out.c_d_id, out.c_w_id, out.d_id, out.w_id, new Date(), out.amount, out.data);
    this.historyDB.insert(history);
}

@Inbound(values = "order-status-out")
@Transactional(type = R)
public void processOrderStatus(OrderStatusOut in){
    Order order = this.orderDB.getLastOrderByCustomerId(in.c_id);
    List<OrderLineInfoDto> orderLinesInfo = this.orderLineDB.getOrderLinesInfo(order.o_id, order.o_d_id, order.o_w_id);
}
\end{lstlisting}
\vspace{-1ex}

\subsection{VMS Bootstrap}
\label{subsec:vms_bootstrap}




Developers implement a VMS through the \name\ software development kit (SDK). Through the SDK, \name\ exposes a set of high-level APIs to enable the specification of VMS constructs, verifies VMS model properties based on the user specification, and pre-loads internal classes to support efficient VMS function execution.

The use of Java is timely as it allows the direct mapping of the programming constructs laid in $\S~\ref{subsec:programming}$ into the SDK. 
Before initializing a VMS instance, the SDK builds a logical representation (referred as \texttt{VMS-Meta}) of the VMS that can be managed by \name\ at runtime based on the VMS specification provided by the user.

\texttt{VMS-Meta} is a set of objects that stores metadata about the data model, data constraints, and the mappings of events to application functions.
These are leveraged to check the correctness of intra-VMS properties, exploring the specification space to identify anomalies related to
functions that append to its own input event log,
event logs with multiple functions appending into it,
functions without input event log,
output events without associated event log, 
writes to foreign data items, 
writes to and reads from non-declared tables,
queries that include non-existing fields,
functions with more than an input event log,
missing data dependence (i.e., incorrect foreign key declaration), 
and more.

These anomalies are detected by navigating through the abstract syntax tree (AST) of user-defined classes and correlating information extracted from the meta-programming constructs, e.g., annotations. If an anomaly is detected, the VMS initialization procedure is stopped and the user is informed about the model violations.


Afterwards, \name\ SDK partners with JVM bootstrap class loader to preload and inject system functionality into key user classes in the execution path of VMS functions so to make their lifecycle and \name\ internals transparent to developers. For instance, it is not a goal to expose how an output event is appended to an event log nor to disclose internal index structures used to store application state.

Specifically, concrete implementations of \texttt{@Repository} interfaces are created via byte code generation and the generated code is instrumented to intercept database operations and forward them to an internal transaction management layer ($\S~\ref{sec:implementation}$). \texttt{@Repository} custom instances are then injected into preloaded \texttt{@Microservice} instances. Besides, complex queries (e.g., with aggregate, join, and where clauses) declared as part of \texttt{@Repository} interfaces 
are parsed and have their query execution plans generated ahead of time to speed up VMS execution.


Apart from managing application entities that map to database rows, 
necessary in every ORM-based application,
no other objects require being managed by users.
The \name\ SDK hides system complexities entirely, 
thereby allowing for clean application logic specification. 


Besides intra-VMS properties, inter-VMS properties also require being checked for correctness. In particular, a single VMS instance appender per event log must hold, event schemas must match across VMS instances, foreign tables from every VMS must map to a single native table from another VMS, unique VMS name identifiers, correctness of transaction definitions, and others.




After initialization, each VMS instance forwards its \texttt{VMS-Meta} to the Catalogs, which are then combined for correctness check of inter-VMS properties by the dispatchers. 
If some property is violated, the VMS instances involved are not allowed to participate in multi-VMS transactions. Afterwards, the \name\ bootstrap process is considered completed.

\subsection{VMS Transaction Scheduling}
\label{subsec:vms_tx_sched}

\subsubsection{Correctness Argument}
We analyze the correctness of the VMS scheduling algorithm in this section. It is based on the assumption that users provide a correct specification of application function execution modes.

\begin{theorem}
Let \( O = [T_1, T_2, \dots, T_n] \) be a total order of transactions defined by the dispatchers, and \( S \) be the schedule produced by the VMS. \( S \) is conflict-equivalent to \( O \).
\end{theorem}

\begin{proof}
The VMS scheduling algorithm ensures that transactions with different execution modes \{Single threaded, Partitioned, Parallel\} are not scheduled concurrently. It puts transactions into a sequence of groups, where all the transactions in a group belong to exactly one of the three execution modes. Transactions in each group have TiDs that are greater than all the TiDs of the transactions in any preceding groups. Furthermore, transactions in each group are completely executed before all the transactions in the subsequent groups, except for read-only (R) transactions. The algorithm ensures that all the R transactions are dispatched after all the potentially conflicting RW transactions with smaller TiDs are executed. Therefore, the multi-version database can ensure the latter reads the version according to the schedule stated in $O$. In the following analysis, we only focus on RW transactions. 

For each transaction group, there are three possible cases:

\fakesub{Case (i)}  All transactions are single-threaded transactions. The VMS scheduling algorithm ensures that for any pair of single-threaded RW transactions, they are executed strictly according to the order in \( O \) without concurrency. 

\fakesub{Case (ii)} All transactions are partitioned transactions. Let \( G(O) \) be the conflict graph of the transactions in this group induced by the total order \( O \), and \( G(S) \) be the corresponding conflict graph induced by the schedule \( S \).

Let \( T_i \to T_j \in G(O) \).  
This implies:
\begin{itemize}
    \item \( T_i \) and \( T_j \) conflict.
    \item \( T_i \prec T_j \) in \( O \).
\end{itemize}
    
The algorithm ensures that RW transactions accessing conflicting partitions are never scheduled concurrently. 
Therefore, for any pair of conflicting transactions \( T_i \) and \( T_j \) such that \( T_i \prec T_j \) in \( O \), the conflicting operations from \( T_i \) precede those from \( T_j \) in \( S \). Hence, the edge \( T_i \to T_j \) exists in the conflict graph \( G(S) \), preserving the conflict-equivalence with \(G(O)\).

\fakesub{Case (iii)} All the transactions are embarrassingly parallel transactions. As we assume that they have no conflicting access to the VMS state, $G(O)$ of the transactions in this group does not have any edges. Therefore, the corresponding $G(S)$ does not violate any precedence relationship in $G(O)$. 

All in all, the complete schedule $S$ is conflict equivalent to the schedule stated by $O$.
\end{proof}

\subsubsection{Detailed Algorithm}
\label{subsubsec:detailed_tx_sched}
Based on Algorithm~\ref{algo:txn_sched_i}, we divide transaction scheduling in a VMS into three functions, \textit{Ingest}, \textit{Schedule}, and \textit{TxnCompletionCallback}, which are executed by different threads concurrently.

The \textit{Ingest} prepares transaction inputs for scheduling by reading all input events available for consumption from the inbound event stream and building transaction contexts for execution. A transaction context maps an input event to the respective application function of a VMS. 
The \textit{Schedule} tracks the execution of transactions continuously, making a best effort to schedule the most amount of concurrent transactions possible, respecting the global order. Like the \textit{Ingest}, it blocks in specific points to release resources for transaction execution. The \textit{TxnCompletionCallback} is executed by the thread assigned to execute a transaction. It updates the \textit{lastTidExec} and the sets \textit{partSched} (set of partitioned transactions scheduled) and \textit{paraSched} (set of parallel transactions scheduled) atomically to allow for the scheduling thread to progress with further transaction scheduling. In case of an error, it triggers the abort procedure, covered in Algorithm~\ref{algo:abort}.

\SetKwProg{Fn}{Function}{:}{end}

\SetKwFunction{FMain}{Schedule}
\SetKwFunction{FIngest}{Ingest}
\SetKwFunction{FProc}{TxnCompletionCallback}

\begin{algorithm*}[t]
\caption{VMS transaction scheduling}
\label{algo:txn_sched_i}
\KwIn{Infinite input event stream $\mathcal{I} = (e_1,e_2,\ldots)$}
\KwIn{$lastTidCommitted$ ($0$ by default)}
\KwOut{Infinite output event stream $\mathcal{O} = (o_1,o_2,\ldots)$}
$txns \gets \emptyset$;\tcp{transaction context set}
$lastTidExec \gets lastTidCommitted$\;
$paraSched \gets \emptyset$;\tcp{parallel txns scheduled}
$partSched \gets \emptyset$;\tcp{partitioned txns scheduled}
\begin{multicols}{2}  
\Fn{\FMain{}}{
    $txn \gets txn' \in txns : \mathrm{txn'.prevTid} = lastTidExec$\;
    \If{$txn.mode = SINGLE$}{
         \If{$\neg(partSched\neq\emptyset \lor paraSched\neq\emptyset) \lor (\exists t:t\in partSched \cup paraSched \land t.type=RW)$}{\label{algoline:singlethread_ok}
            \textbf{block while} above cond. holds\; \label{algoline:singlethread_block}
        }
        \lIf{t.type=R}{\textbf{dispatch}$(t)$;
        \textbf{else} {{\bf execute}$(t)$}}   \label{algoline:singlethread_end} 
        \textbf{goto line 6}\;
    }
    \Do{$txn \ne \emptyset \land execModeAux = txn.mode$}{
        \uIf{$txn.mode = PARA \land (partSched=\emptyset \lor (\forall t:t\in partSched \land t.type=R))$}{
            $paraSched \gets paraSched \cup txn$\;
            \textbf{dispatch}$(txn)$\;
        }
        \uElseIf{$txn.mode = PART \land (paraSched=\emptyset \lor(\forall t:t\in paraSched \land t.type=R))$}{
            \uIf{$partSched[t.part] = \emptyset \lor (\forall t: t\in partSched[t.part]\land t.type=R$)}{
             $partSched \gets partSched \cup txn$\;
                \textbf{dispatch}$(txn)$\;
            }\lElse{ \textbf{block until} $partSched[t.part] = \emptyset$
            }
        }
        \Else{
            \textbf{goto line 6}\;
        }
        $nextTxn \gets txn' \in txns : \mathrm{txn'.prevTid} = txn.Tid$\;
        $execModeAux \gets txn.mode$\;
        $txn \gets nextTxn$\;
    }
}
\columnbreak   
\Fn{\FIngest{}}{
    \While{true}{
        $I' \gets$ \textbf{pollAllBlocking}$(\mathcal{I})$ where $I' \subset I$\;
        \For{$e \in \mathcal{I'}$}{
            $f \gets$ \textbf{mapVmsFunction}$(e)$\;
            $txnCtx \gets$ \textbf{buildTxnCtx}$(e,f)$\;
            $txns \gets txns \cup txCtx$
        }
    }
}
\Fn{\FProc{$txn, status$}}{
    \If{$status = ERROR$}{
        \textbf{abort}$(txn)$\;
        \textbf{return}\;
    }
    $lastTidExec \gets max(lastTidExec, txn.Tid)$\;
    $txn.isFinished \gets true$\;
    $\mathcal{O} \gets \mathcal{O} \cup txn.result$\;
    \uIf{txn.mode = PARA}{
        $paraSched \gets paraSched \setminus txn$\;
    }\ElseIf{txn.mode = PART}{
        $partSched \gets partSched \setminus txn$\;
    }
}
\end{multicols}
\end{algorithm*}
 
\SetKw{KwBlock}{block}
\SetKw{KwUnblock}{unblock}

\begin{algorithm}[tb]
\caption{VMS transaction abort}
\label{algo:abort}
\KwIn{$txn$ that led to a conflict or constraint violation}
\KwIn{Infinite input event stream $\mathcal{I} = (e_1,e_2,\ldots)$}
\KwIn{Transaction context set $txns$}
\KwIn{Set of VMS tables $T$}
\KwBlock{($I$)};\tcp{block \textit{Ingest} polling $I$}
\KwBlock{($txns$)};\tcp{block \textit{Schedule} access to $txns$}
\If{($txn$ abort is raised by this VMS)}{
    \textbf{forward abort msg to COMMIT\_HANDLER}\;
}
$txnsToAbort \gets \forall txn' \in txns : \mathrm{txn'.Tid} \geq txn.Tid$\;
$txns \gets txns \setminus txnsToAbort$\;
\For{$table \in T$}{
    \For{$txnToAbort \in txnsToAbort$}{
        $table.versions \gets table.versions \setminus txnToAbort.writeSet$\;
    }
}
\uIf{(VMS is the first node in the transaction graph)}{
    log $txn$ abort event in VMS state\;
    find the successor and predecessor events of $txn$ aborted in I\;
    adjust dependencies\;
    reinsert events after $txn$ aborted in $I$\;
}\Else {
    find the predecessor event of $txn$ aborted in $I$\;
    wait until successor event of the predecessor is sent through $I$\;
}
$lastTidExec \gets predecessor.Tid$\;
\KwUnblock{($I$)}\;
\KwUnblock{($txns$)};
\end{algorithm}

\subsection{Fault Tolerance}
\label{subsec:fault_tolerance}

Assume that \name\ components can crash at any time, resulting in the proxies, dispatchers, commit handlers, and VMSes losing their data in memory. 

\noindent\textbf{VMSes.}
The VMSes can replicate the event logs to standby replicas for failure recovery. In this case, sender I/O threads, besides logging log segments, can stream them to replicas. The replicas process the log segments in the same way as the primary but do not send output events to consumer VMSes.
For failure detection, replicas rely on heartbeats received from primary nodes. Upon a detected failure, replicas trigger a leader election. Similarly to Raft~\cite{raft}, replicas start as candidates and rely on the number of votes received to decide whether the candidacy is successful. In this case, it broadcasts the election result. As transactions execute optimistically ($\S$~\ref{subsec:tx_mgmt}), a newly elected leader may not contain some log segments. Therefore, it is necessary to request possible missing log segments from upstream VMSes. Note that the Catalogs must also be checked for abort information of in-progress batches. 
In this case, replicas must proceed in the same way as the primary to safeguard correctness. In other words, Algo.~\ref{algo:abort} must also be executed.
Thereafter, the system can return to normal operation.





\noindent\textbf{Proxies.} Proxies act as stateless load balancers. In case of failures or newly deployed proxies, they obtain from the Catalogs the latest VMS metadata and transaction definition versions stored. This procedure repeats if an invocation fails due to a deprecated VMS metadata or transaction definition. 

\noindent\textbf{Dispatchers and Commit Handlers.}
In this algorithm, we assume both dispatchers and commit handlers execute through multiple threads within a process so to use multi-core machines. We also assume the Catalogs is maintained as a shared data structure within the same process. We leave scaling out to multi-node through multiple processes to future work.

To make the commit handling fault-tolerant, upon all $batch\_complete$ messages' arrival, a commit handler logs a $batch\_commit$ message to the Catalogs. 
Upon a crash, a stand-by replica requests the last committed batch ID from all VMSes to cover cases where the commit handler crashes before logging it. Upon receiving the responses, it updates its last committed batch ID if necessary and sends a $batch\_aborted$ message to all VMSes to abort all ongoing batches at the time of the crash. VMSes then proceed similarly to a transaction abort. VMSes discard non-committed data item versions, and TiD precedences are adjusted accordingly to receive transactions from the new process. 

Differently from VMSes, the dispatchers do not stream event logs to replicas since these are simply transaction requests that can be retried by clients. Although this may lead to losing some useful work the VMSes performed, it avoids a TCP round trip to each replica for every transaction emitted.

\subsection{VMS Execution Modes}
\label{app:subsec:concurrency}

Table~\ref{tab:vms_concurrency_om} exhibits the VMS execution modes used in Online Marketplace and Table~\ref{tab:vms_concurrency_tpcc} exhibits for TPC-C. In Dapr, all operations run as a parallel task given the lack of transaction isolation.

\begin{table}[tb]
\centering
\caption{VMS execution modes in Online Marketplace}
\vspace{-2ex}
\small
\begin{tabularx}{\columnwidth}{|X|X|X|}
\hline
\textbf{Component} & \textbf{Transaction} & \textbf{Primitive} \\
\hline
Cart & Customer Checkout & Keyed on \texttt{customer\_id} \\
\hline
Cart & Price Update & Single thread; can't extract from the event which carts contain the product \\
\hline
Stock & Customer Checkout & Runs single-threaded in the experiments but can be keyed on \texttt{seller\_id, product\_id} \\
\hline
Product, Cart, Stock & Product Delete & Keyed on \texttt{seller\_id, product\_id} \\
\hline
Product & Price Update & Keyed on \texttt{seller\_id, product\_id} \\
\hline
Order & Customer Checkout & Keyed on \texttt{customer\_id} \\
\hline
Payment & Customer Checkout & Parallel \\
\hline
Shipment & Customer Checkout & Parallel \\
\hline
Seller & Seller Dashboard & Parallel \\
\hline
\end{tabularx}
\vspace{-1ex}
\label{tab:vms_concurrency_om}
\vspace{-1ex}
\end{table}

\begin{table}[tb]
\centering
\caption{VMS execution modes in TPC-C}
\vspace{-2ex}
\small
\begin{tabularx}{\columnwidth}{|X|X|X|X|}
\hline
\textbf{Component} & \textbf{New Order} & \textbf{Payment} & \textbf{Order Status} \\
\hline
Warehouse & Keyed on warehouse and district IDs & Keyed on warehouse and district IDs & Read-only, thus parallel \\
\hline
Inventory & Runs keyed on supplier warehouse IDs in experiments but can be keyed on item IDs for increased concurrency & - & - \\
\hline
Order & Parallel & Parallel & Read-only, thus parallel \\
\hline
\end{tabularx}
\vspace{-1ex}
\label{tab:vms_concurrency_tpcc}
\vspace{-1ex}
\end{table}

\subsection{PostgreSQL Tuning}
\label{app:subsec:pg_tuning}

In the application ORM, we set the maximum number of connections to 10K, bypassing the default maximum of 100 connections, and we configure a connection pool to reuse connections. 
Besides, we disable synchronous commit, so updates are eventually logged rather than atomically\footnote{In off mode, there is no waiting, so there can be a delay between when success is reported to the client and when the transaction is guaranteed to be safe against a server crash (the maximum delay is 3X $wal\_writer\_delay$).}
We use Unix domain socket for inter-process, kernel-based communication instead of the default TCP sockets.
We increase the pgsql buffer size to improve analytical query performance.
We increase the shared buffer from 128MB to 3GB, which corresponds to \~15\% of 21 GB available in c5n.2xlarge instance, as suggested by PostgreSQL documentation.
The work mem is set to 128 MB instead of the default 4MB.

\subsection{Effect of Batch Size}
To study the effects of the number of transactions per batch in OM, we start by fixing the batch epoch to infinity, varying the number of transactions in a batch, and maximizing the concurrency level. In Figure~\ref{fig:batch_size_a} (left), we observe an average of 1K tx/sec increase up to batch size 1K, the point where throughput stabilizes. Next, we fix the maximum number of transactions in a batch to infinity and vary the epoch. Figure~\ref{fig:batch_size_a} (right) demonstrates the effectiveness of the commit protocol, showing little variation across epochs.


\begin{figure}[H]
\captionsetup[subfigure]{labelformat=empty}
\centering
\subfloat[]{
  \input{plots/06c_batch}
}
\subfloat[]{
  \input{plots/07c_batch}
}
\vspace{-6ex}
\caption{Effect of Batch Size on Throughput}
\label{fig:batch_size_a}
\vspace{-2ex}
\end{figure}

Figure~\ref{fig:batch_size_b} (left) shows the substantial overhead of network I/O if using a single transaction per batch. It is worth noting that 10 transactions per batch achieves low latency and doubles Dapr's throughput, providing a good trade-off in this particular scenario. 
Besides, Figure~\ref{fig:batch_size_b} (right) shows the insensitivity of \texttt{Seller Dashboard}'s latency to the batch sizes. On the other hand, Figure~\ref{fig:batch_size_c} shows that varying the batch epoch leads to a more natural variation in latency for transactions committing in a batch, without affecting throughput though.

\begin{figure}[tp]
\centering
  \input{plots/06a_batch}
   \input{plots/06b_batch}
\caption{Effect of Batch Size on Latency}
\label{fig:batch_size_b}
\end{figure}

\begin{figure}[tp]
\centering
 \input{plots/07a_batch}
  \input{plots/07b_batch}
\caption{Effect of Batch Epoch on Latency}
\label{fig:batch_size_c}
\end{figure}


\subsection{Effect of Scale Factor on Latency}

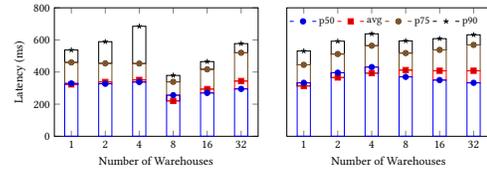
\begin{figure}[H]
\centering
    \input{plots/09_batch_latency_mem_tpcc}
    \input{plots/09_batch_latency_disk_tpcc}
\vspace{-2ex}
\caption{Effect of Scale Factor on Batch Latency in TPC-C. In-memory execution (left) and with durability (right)}
\label{fig:scale_factor_effect}
\vspace{-3ex}
\end{figure}

Figure~\ref{fig:scale_factor_effect} reports the statistics of the intervals between the commit of each pair of adjacent batches for each scale factor in TPC-C. In \name, batch latencies decrease because batches are filled and processed more quickly due to the higher number of requests. This effect is also observable in \name\ with logging and checkpointing with an average of 17\% overhead due to the intensive I/O incurred by the durable write procedures. 


%% file: plots/06c_batch.tex
\begin{tikzpicture}[scale=0.5]
\begin{axis}[
  width=7cm,
  height=3.5cm,
        ybar,
        xlabel={Number of Transactions per Batch},
        ylabel=Throughput (tx/s),
        xtick=data,
        xticklabels={1,10,100,1K,10K},
        ymin=0
        ]
        \addplot
            coordinates {
            (1,11388)(2,12900)(3,13612)(4,14399)(5,14465)
            };
    \end{axis}
\end{tikzpicture}

%% file: plots/07c_batch.tex
\begin{tikzpicture}[scale=0.5]
\begin{axis}[
        width=7cm,
        height=3.5cm,
        ybar,
        xlabel={Batch Epoch (ms)},
        ytick=\empty,
        xtick=data,
        xticklabels={10,50,100,200,500},
        ymin=0
        ]
        \addplot
            coordinates {
            (0,13369)(1,13931)(2,14554)(3,14432)(4,14465)
            };
    \end{axis}
\end{tikzpicture}

%% file: plots/06a_batch.tex
\begin{tikzpicture}[scale=0.5]
\begin{axis}[
  width=7cm,height=4.85cm,
        stack plots=y,
        /tikz/ybar,
        ymin=0,
        xlabel={Num Txns/Batch w/o Seller Dashboard},
        ylabel=End-to-end Latency (ms),
        xtick=data,
        xticklabels={1,10,100,1K,10K},
        legend style={legend columns=1,fill=none,draw=none,
        legend pos=north east,font=\normalsize
        }
        ]
        \addplot
            coordinates {
            (1,2975)(2,0.79)(3,265)(4,268)(5,272.5)
            };
        \addplot
            coordinates {
            (1,375)(2,4.14)(3,8)(4,2)(5,0.25)
            };
        \addplot
            coordinates {
            (1,2079)(2,1.11)(3,137)(4,129)(5,117.25)
            }; 
        \addplot
            coordinates {
            (1,1656)(2,9.81)(3,95)(4,88)(5,103)
            };
        \legend{p50,avg,p75,p90}
    \end{axis}
\end{tikzpicture}

%% file: plots/06b_batch.tex
\begin{tikzpicture}[scale=0.5]
\begin{axis}[
  width=7cm,height=4.85cm,
        stack plots=y,
        /tikz/ybar,
        ymin=0,
        xlabel={Num Txns/Batch only Seller Dashboard},
        xtick=data,
        xticklabels={1,10,100,1K,10K},
        legend style={legend columns=4,fill=none,draw=none,
        legend pos=north east,font=\small
        }
        ]
        \addplot
            coordinates {
            (1,2.47)(2,1.58)(3,1.3)(4,1.57)(5,1.47)
            };
        \addplot
            coordinates {
            (1,7.84)(2,5.42)(3,6.45)(4,3.72)(5,3.58)
            };
        \addplot
            coordinates {
            (1,-7.14)(2,-3.88)(3,-4.16)(4,-1.15)(5,-1.58)
            }; 
        \addplot
            coordinates {
            (1,10.13)(2,8.68)(3,8.71)(4,8.56)(5,8.38)
            };
    \end{axis}
\end{tikzpicture}

%% file: plots/07a_batch.tex
\begin{tikzpicture}[scale=0.5]
\begin{axis}[
        width=7cm,height=4.85cm,
        stack plots=y,
        /tikz/ybar,
        ymin=0,
        xlabel={Batch Epoch (ms) w/o Seller Dashboard},
        ylabel=End-to-end Latency (ms),
        xtick=data,
        xticklabels={10,50,100,200,500},
        legend style={legend columns=1,fill=none,draw=none,
        legend pos=north west,font=\normalsize}
        ]
        \addplot
            coordinates {
            (0,8.18)(1,28)(2,52)(3,102)(4,272.5)
            };
        \addplot
            coordinates {
             (0,3.45)(1,4)(2,6)(3,10)(4,0.25)
            };
        \addplot
            coordinates {
             (0,4.31)(1,13)(2,17)(3,53)(4,117.25)
            }; 
        \addplot
            coordinates {
             (0,11.11)(1,19)(2,26)(3,49)(4,103)
            };
        \legend{p50,avg,p75,p90}
    \end{axis}
\end{tikzpicture}

%% file: plots/07b_batch.tex
\begin{tikzpicture}[scale=0.5]
\begin{axis}[
        width=7cm,height=4.85cm,
        stack plots=y,
        /tikz/ybar,
        ymin=0,
        xlabel={Batch Epoch (ms) only Seller Dashboard},
        xtick=data,
        xticklabels={10,50,100,200,500},
        legend style={legend columns=4,fill=none,draw=none,
        legend pos=north east,font=\small}
        ]
         \addplot
            coordinates {
            (0,1.64)(1,1.69)(2,1.76)(3,1.65)(4,1.47)
            };
        \addplot
            coordinates {
            (0,5.82)(1,5.16)(2,4.01)(3,3.86)(4,3.58)
            };
        \addplot
            coordinates {
            (0,-3.28)(1,-2.53)(2,-0.61)(3,-1.13)(4,-1.58)
            }; 
        \addplot
            coordinates {
            (0,9.15)(1,9.03)(2,8.59)(3,8.93)(4,8.38)
            };
    \end{axis}
\end{tikzpicture}

%% file: plots/09_batch_latency_mem_tpcc.tex
\begin{tikzpicture}[scale=0.5]
\begin{axis}[
        width=7cm,height=5cm,
        stack plots=y,
        /tikz/ybar,
        ymin=0,
        xlabel={Number of Warehouses},
        ylabel=Latency (ms),
        xtick=data,
        xticklabels={1,2,4,8,16,32},
        legend style={legend columns=4,fill=none,draw=none,
        legend pos=north east,font=\small}
        ]
        \addplot
            coordinates {
            (1,398)(2,289)(3,174)(4,121)(5,88)(6,83)
            };
        \addplot
            coordinates {
            (1,6)(2,9)(3,12)(4,4)(5,27)(6,19)
            };
        \addplot
            coordinates {
            (1,16)(2,35)(3,36)(4,30)(5,18)(6,10)
            }; 
        \addplot
            coordinates {
            (1,8)(2,14)(3,31)(4,48)(5,76)(6,52)
            };
    \end{axis}
\end{tikzpicture}

%% file: plots/09_batch_latency_disk_tpcc.tex
\begin{tikzpicture}[scale=0.5]
\begin{axis}[
        width=7cm,height=5cm,
        stack plots=y,
        /tikz/ybar,
        ymin=0,
        xlabel={Number of Warehouses},
        ylabel={\phantom{invisible}},
        xtick=data,
        xticklabels={1,2,4,8,16,32},
        legend style={legend columns=2,fill=none,draw=none,
        legend pos=north east}
        ]
        \addplot
            coordinates {
            (1,476)(2,369.5)(3,239)(4,137)(5,115)(6,88)
            };
        \addplot
            coordinates {
              (1,-2)(2,0.2)(3,-4)(4,20)(5,17)(6,31)
            };
        \addplot
            coordinates {
              (1,16)(2,23)(3,37)(4,42)(5,50)(6,52)
            }; 
        \addplot
            coordinates {
             (1,19)(2,30)(3,37)(4,73)(5,41)(6,78)
            };
        \legend{p50,avg,p75,p90}
    \end{axis}
\end{tikzpicture}